\documentclass[draft]{agujournal2019}
\usepackage{url, amsmath, amsfonts, booktabs} 
\usepackage{subfiles}
\usepackage{lineno}
\usepackage[inline]{trackchanges} 
\usepackage{soul}

\newcommand*\patchAmsMathEnvironmentForLineno[1]{
  \expandafter\let\csname old#1\expandafter\endcsname\csname #1\endcsname
  \expandafter\let\csname oldend#1\expandafter\endcsname\csname end#1\endcsname
  \renewenvironment{#1}
  {\linenomath\csname old#1\endcsname}
  {\csname oldend#1\endcsname\endlinenomath}}
  \newcommand*\patchBothAmsMathEnvironmentsForLineno[1]{
  \patchAmsMathEnvironmentForLineno{#1}
  \patchAmsMathEnvironmentForLineno{#1*}}
  \AtBeginDocument{
  \patchBothAmsMathEnvironmentsForLineno{equation}
  \patchBothAmsMathEnvironmentsForLineno{align}
  \patchBothAmsMathEnvironmentsForLineno{flalign}
  \patchBothAmsMathEnvironmentsForLineno{alignat}
  \patchBothAmsMathEnvironmentsForLineno{gather}
  \patchBothAmsMathEnvironmentsForLineno{multline}
}

\draftfalse

\journalname{Geophysical Research Letters}

\begin{document}

%
%



\title{Generalizable neural-network parameterization of mesoscale eddies in idealized and global ocean models}

%
%



\authors{Pavel Perezhogin\affil{1}, Alistair Adcroft\affil{3}, Laure Zanna\affil{1,2}}


\affiliation{1}{Courant Institute of Mathematical Sciences, New York University, New York, NY, USA}
\affiliation{2}{Center for Data Science, New York University, New York, NY, USA}

\affiliation{3}{Program in Atmospheric and Oceanic Sciences, Princeton University, Princeton, NJ, USA}




\correspondingauthor{Pavel Perezhogin}{pp2681@nyu.edu}




\begin{keypoints}

\item Physics constraints are developed for a neural-network parameterization of mesoscale eddy fluxes

\item Dimensional scaling constraints improve offline generalization to unseen grid resolutions and depths


\item New parameterization improves the representation of kinetic and potential energy online in coarse idealized and global ocean models


\end{keypoints}


%
%

%
%


\begin{abstract}
Data-driven methods have become popular to parameterize the effects of mesoscale eddies in ocean models. However, they perform poorly in generalization tasks and may require retuning if the grid resolution or ocean configuration changes. We address the generalization problem by enforcing physics constraints on a neural network parameterization of mesoscale eddy fluxes. We found that the local scaling of input and output features helps to generalize to unseen grid resolutions and depths offline in the global ocean. The scaling is based on dimensional analysis and incorporates grid spacing as a length scale. We formulate our findings as a general algorithm that can be used to enforce data-driven parameterizations with dimensional scaling. The new parameterization improves the representation of kinetic and potential energy in online simulations with idealized and global ocean models. Comparison to baseline parameterizations and impact on global ocean biases are discussed. 
\end{abstract}

\section*{Plain Language Summary}
Ocean models can't directly simulate eddies that are smaller than the resolution of the computational grid. The effect of these eddies is represented by parameterizations.  Machine learning offers a new way to build parameterizations directly from data, however, such parameterizations may fail when tested in new, unseen scenarios. Here, we leverage physics constraints to mitigate this, generalization, problem. Specifically, we found that method of dimensional analysis can be used to constrain data-driven parameterizations to enhance their accuracy in new scenarios without the need for retraining. New parameterization is tested in a realistic ocean model and brings us closer to robust, data-driven methods for ocean and climate models.

%
%

%


%
%
%
%

\section{Introduction}
Numerical ocean models rely on parameterizations to represent the effects of physical processes smaller than the model grid spacing, which are unresolved \cite{fox2019challenges, hewitt2020resolving, christensen2022parametrization}. 
Recently, there has been a growing interest in applying machine learning methods to parameterize these subgrid physics in ocean models \cite{Bolton2019, zanna2020data, guillaumin2021stochastic, zhang2023implementation, Sane2023, yan2024choice, perezhogin2024stable, maddison2024online}. However, developing data-driven parameterizations for ocean models is still in its early stages, and their application is often limited to idealized configurations. Deploying data-driven parameterizations in the global ocean presents several challenges, one of which is addressed in this study -- the problem of generalization to unseen scenarios.

Data-driven parameterizations rely heavily on sets of training data, and their successful implementation often requires tuning when applied to a new grid resolution \cite{zhang2023implementation}, flow regime \cite{ross2022benchmarking}, model configuration \cite{perezhogin2024stable}, depth, or geographical region \cite{gultekin2024analysis}. However, in practice, it would be desirable to have a single parameterization that performs effectively across a variety of scenarios without requiring retuning. The ability of a data-driven model to work on new (testing) data, which is distinct from the training data, is measured by the \textit{generalization error} \cite{bishop2006pattern, hastie2009elements}. Data-driven methods work best when the testing data is drawn from the same distribution as the training data. However, in geophysical applications, the distribution of physical variables can vary vastly across different scenarios—a phenomenon referred to as a \textit{distribution shift} \cite{beucler2024climate, gultekin2024analysis}. In this case, domain knowledge and physics constraints can be leveraged to mitigate the generalization error of data-driven models \cite{kashinath2021physics}.

In this work, we demonstrate how physics constraints can be leveraged to enhance the generalization of an Artificial Neural Network (ANN) parameterization of the ocean mesoscale eddy fluxes. Following \citeA{beucler2024climate}, we rescale features of the ANN to minimize the distribution shift. To identify a suitable normalization technique for eddy fluxes, we apply dimensional analysis and \citeA{buckingham1914physically}'s Pi theorem. Specifically, we introduce a local dimensional scaling constructed from the grid spacing and velocity gradients \cite{prakash2022invariant}. The local scaling improves offline generalization of the ANN parameterization to unseen grid resolutions and depths, as found in the global ocean dataset CM2.6 \cite{griffies2015impacts}. Our findings are formulated as a general algorithm that can be used to incorporate the dimensional scaling in future applications. Additional physics constraints for the ANN parameterization are enforced following \citeA{guan2022learning} and \citeA{srinivasan2024turbulence}. 
We present an online evaluation of the new ANN parameterization in the GFDL MOM6 ocean model \cite{adcroft2019gfdl} in idealized and global configurations.

\section{A Method to Constrain Neural Network with Dimensional Scaling} \label{sec:unit_invariance}

Here we introduce the concept of physical dimensionality and demonstrate how it can be used to constrain data-driven parameterizations. 
We start with a trivial example, followed by a general algorithm. Finally, we draw connections to existing approaches.


\subsection{Trivial Example}

Consider the case where a scalar momentum flux $T$ (units of $\mathrm{m}^2 \mathrm{s}^{-2}$) can be predicted using a length scale $\Delta$ (units of $\mathrm{m}$) and inverse  time scale $X$ (units of $\mathrm{s}^{-1}$):
\begin{equation}
    T = f(\Delta, X). \label{eq:eq_dimensional}
\end{equation}
Eq. \eqref{eq:eq_dimensional} must remain invariant under rescaling the units of time and length, that is for any $\alpha, \beta>0$, the equality must hold: $f( \alpha \Delta, \beta X) = \alpha^2 \beta^2 f(\Delta, X)$. However, the unit invariance can be violated when $f$ is parameterized by neural networks. One way to enforce it is by leveraging \citeA{buckingham1914physically}'s Pi theorem, which states that the dimensional equation  (such as Eq. \eqref{eq:eq_dimensional}) can be rewritten in non-dimensional form. Specifically, for a set of three dimensional variables ($T$, $\Delta$, $X$) with two independent dimensions (length and time), there is only one (three minus two) non-dimensional variable ($\pi_1=T/ (\Delta^2 X^2)$). Thus, Eq. \eqref{eq:eq_dimensional} transforms to  $\pi_1=\mathrm{const}$, or equivalently: \begin{equation}
    T = \Delta^2 X^2 \theta, \label{eq:non_dim}
\end{equation}
where $\theta$ can be interpreted as a non-dimensional \citeA{smagorinsky1963general} coefficient. A data-driven parameterization in the form of Eq. \eqref{eq:non_dim} with a trainable parameter $\theta$, which is constant, follows the dimensional scaling as a hard constraint, in contrast to Eq. \eqref{eq:eq_dimensional}, which does not guarantee dimensional consistency. Eq. \eqref{eq:non_dim} promotes generalization as it explicitly accounts for the change in the magnitude of independent variables ($\Delta$ and $X$), constraining the learnable part of the mapping ($\theta$) to be on the order of unity. 

\subsection{General Algorithm} \label{sec:general_algorithm}

Extending the example above, we suggest an algorithm to enforce dimensional scaling in  ANN parameterizations by preprocessing input and output features:
\begin{enumerate}
    \item Identify the input features that contribute significantly to the accurate prediction of the output features;
    \item Construct non-dimensional input and output features from a \textit{combined} set of identified input and output features;
    \item Verify that a traditional known parameterization is a special case of the constructed non-dimensional mapping. 
\end{enumerate}

Step 1 follows standard dimensional analysis textbooks \cite{bridgman1922dimensional, barenblatt1996scaling}.
Specifically, a relevant set of input features can be identified by physical intuition or through ablation studies by evaluating the gain in offline performance from including additional dimensional features in the input set. 
Constructing non-dimensional features is a common approach in physics-constrained data-driven parameterizations \cite{ling2016reynolds, schneider2024opinion}. However, the normalization of input features is often considered separately from the normalization of output features \cite{xie2020modeling, kang2023neural, beucler2024climate, christopoulos2024online}, unlike our proposed method (step 2 above). Additionally, the emphasis in these works is often placed on identifying normalization factors that minimize the distribution shift, while we suggest starting with identifying features responsible for the prediction (step 1). Finally, traditional parameterizations are often used to propose efficient normalization factors \cite{xie2020modeling, kang2023neural, connolly2025deep}, while we instead advocate for having traditional parameterizations as a special case (step 3, \citeA{prakash2022invariant, prakash2024invariant}).

\section{Physics Constraints for Ocean Mesoscale Parameterization}

\begin{figure}[h!]
\centering{\includegraphics[width=0.9\textwidth]{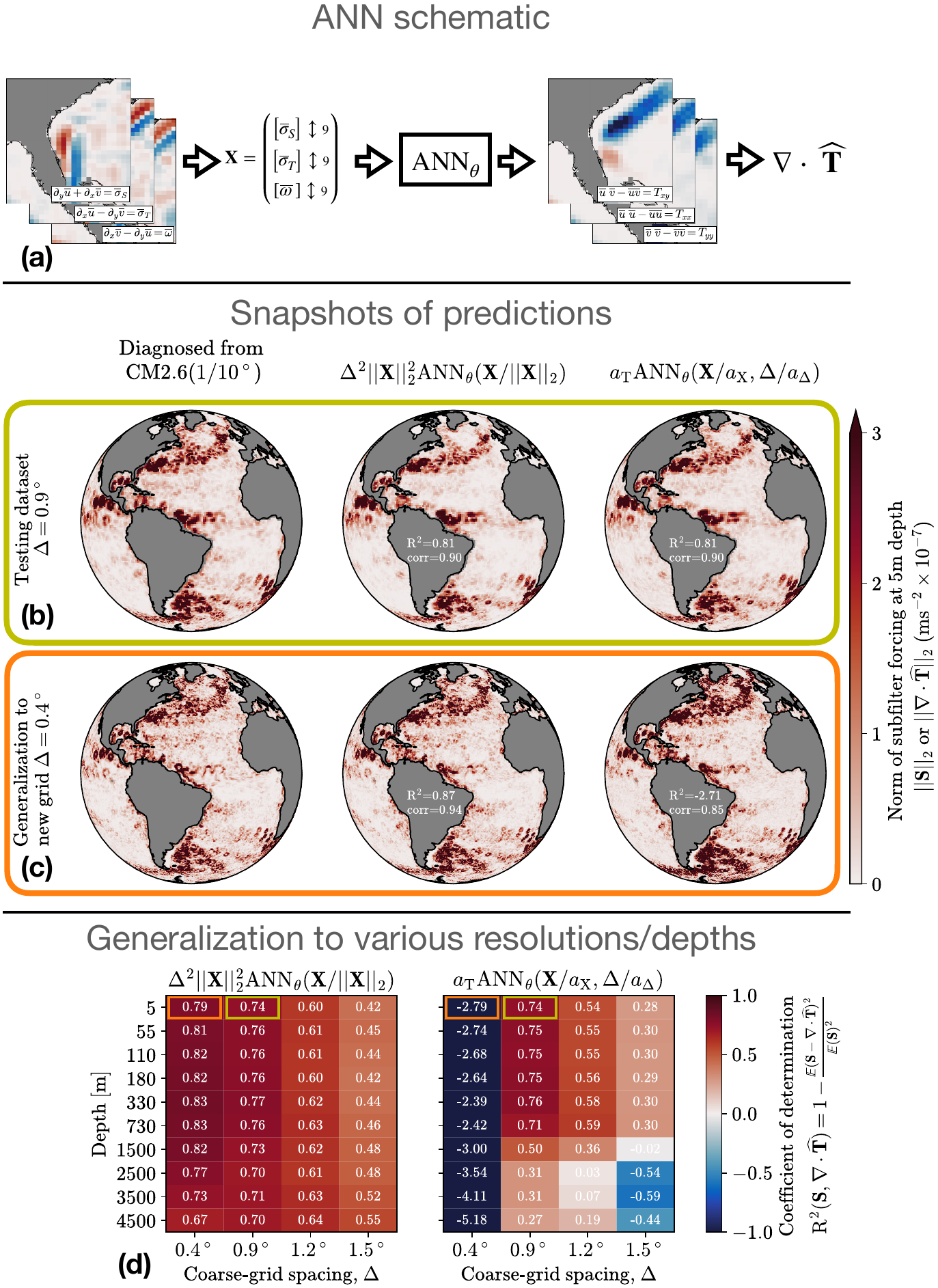}}
\caption{(a) Artificial neural network (ANN) parameterization predicting the divergence of subfilter fluxes given the velocity gradients on the horizontal stencil of $3$$\times$$3$ points. (b) Snapshots of predictions by two ANNs: with local dimensional scaling (Eq. \eqref{eq:model_3}, center column) or with fixed normalization coefficients (Eq. \eqref{eq:model_1}, right column) at the resolution ($0.9^\circ$) and depth ($5\mathrm{m}$) used for training (testing data is separated by 10 years). (c) Prediction at the unseen resolution ($0.4^\circ$) and the same depth ($5$m). 
(d) Coefficient of determination ($\mathrm{R}^2$) in prediction of subfilter forcing for various resolutions and depths, different from that used for training ($0.9^\circ$, $5$m). The $\mathrm{R}^2$ is averaged over 2 years of held-out data and excludes 2 grid points adjacent to the coastline, where green and orange boxes correspond to panels (b) and (c), respectively.
}
\label{fig:offline}
\end{figure}

Our goal is to predict the subfilter momentum fluxes of mesoscale eddies using an Artificial Neural Network (ANN) parameterization, see schematic in Figure \ref{fig:offline}(a). Various physical invariances were imposed to promote generalization.

\subsection{Learning Subfilter Fluxes}
We consider the acceleration produced by subfilter ocean mesoscale eddies \cite<subfilter forcing,>{Bolton2019}:
\begin{equation}
    \partial_t \overline{\mathbf{u}} = \mathbf{S} = (\overline{\mathbf{u}} \cdot \nabla ) \overline{\mathbf{u}} - \overline{(\mathbf{u} \cdot \nabla) \mathbf{u}},  \label{eq:subfilter_forcing}
\end{equation}
where $\mathbf{u}=(u,v)$ is the horizontal ocean  velocity, $\nabla=(\partial_x, \partial_y)$ is the horizontal gradient, and $\overline{(\cdot)}$ is the horizontal filter. The subfilter forcing can be approximated \cite{loose2023comparing} as a divergence of the momentum flux:
\begin{equation}
    \mathbf{S} \approx \nabla \cdot \mathbf{T} \label{eq:momentum_div}
\end{equation}
where
\begin{equation}
    \mathbf{T} = \begin{pmatrix}
        T_{xx} & T_{xy} \\
        T_{yx} & T_{yy}
    \end{pmatrix} = \begin{pmatrix}
        \overline{u}  \, \overline{u} - \overline{u u} & \overline{u} \, \overline{v} - \overline{uv} \\
        \overline{u} \, \overline{v} - \overline{uv} & 
        \overline{v}  \, \overline{v} - \overline{v v}
    \end{pmatrix}. \label{eq:momentum_flux}
\end{equation}
We predict the three components of $ \mathbf{T}$, namely   $T_{xx}$, $T_{xy}$, $T_{yy}$, rather than $\mathbf{S}$ directly, similarly to \citeA{zanna2020data} (ZB20 hereafter) to impose momentum conservation as a hard constraint. We enforce symmetry of the tensor $\mathbf{T}$ by predicting $T_{xy}$ and sharing its prediction with $T_{yx}$, which guarantees angular momentum conservation \cite[Section 17.3.3]{griffies2018fundamentals}, up to machine precision. We also promote rotational and reflection invariances via data augmentation \cite{guan2022learning}, independently rotating each training snapshot by $90^\circ$ and reflecting it along the $x$ and $y$ axes, resulting in $8=2^3$ augmented snapshots per original one.

We learn the components of $\mathbf{T}$ by minimizing the  mean squared error (MSE) loss function:
\begin{equation}
\mathcal{L}_{\mathrm{MSE}} = ||(\mathbf{S} - \nabla \cdot \widehat{\mathbf{T}}) \cdot m ||_2^2 ~ / ~||\mathbf{S} \cdot m||_2^2, \label{eq:mse_loss}
\end{equation}
where $m$ is the mask of wet points and $\widehat{\mathbf{T}}$ is the neural network prediction of the subfilter flux as discussed below. See SI for further details.  

\subsection{Input Features}
Here, we identify the input features relevant for the prediction of momentum fluxes (step 1 of the algorithm presented in Section \ref{sec:general_algorithm}).

Following ZB20, we consider the components of the strain-rate tensor and vorticity as input features:
\begin{gather}
    \begin{aligned}
        \overline{\sigma}_S & = \partial_y \overline{u} + \partial_x \overline{v} && \text{-- shearing strain,} \\
        \overline{\sigma}_T & = \partial_x \overline{u} - \partial_y \overline{v} && \text{-- horizontal tension/stretch,} \\
        \overline{\omega} & = \partial_x \overline{v} - \partial_y \overline{u} && \text{-- relative vorticity.}
    \end{aligned} \label{eq:ZB_features}
\end{gather}
These input features exclude explicit dependence on the velocity, guaranteeing Galilean invariance of the parameterization \cite{srinivasan2024turbulence,pope1975more, lund1993parameterization, ling2016reynolds}. When using the input features (Eq. \eqref{eq:ZB_features}) pointwise, the resulting ANN parameterization highly correlates with the ZB20 equation-discovery model. Thus, we decided to include the non-local contribution of these features \cite{srinivasan2024turbulence, maulik2017neural, maulik2019subgrid, pawar2020priori, wang2021artificial, wang2022constant, gultekin2024analysis}. To do so, the input vector to the ANN consists of velocity gradients, each defined on a $3 \times 3$ horizontal stencil and flattened into a vector of length 9 (denoted as $\left[ \cdot \right]{\updownarrow \scriptstyle 9}$):
\begin{equation}
    \mathbf{X} =
    \begin{pmatrix}
        \left[ \overline{\sigma}_S \right]{\updownarrow \scriptstyle 9}  \\
        \left[ \overline{\sigma}_T \right]{\updownarrow \scriptstyle 9} \\
        \left[ \overline{\omega} \right]{\updownarrow \scriptstyle 9}
    \end{pmatrix} \in \mathbb{R}^{27}. \label{eq:stack_features}
\end{equation}

To facilitate generalization across different resolutions \cite<scale-aware or grid-aware parameterization,>{bachman2017scale}, we account for the local grid spacing of the coarse resolution model $\Delta = \sqrt{\Delta x \Delta y}$, resulting in the following functional form of the parameterization \cite{lund1993parameterization, li2025transformer}:
\begin{equation}
    \mathbf{T} \approx \widehat{\mathbf{T}}(\mathbf{X}, \Delta). \label{eq:model_2}
\end{equation}
Accounting for grid spacing is physically justified as velocity gradients and momentum fluxes differ in dimensionality and require a length scale to be invoked.

\subsection{Neural Network Parameterizations} \label{sec: eddy_param}

We consider a baseline data-driven parameterization of eddy fluxes with fixed normalization coefficients, following the form of Eq. \eqref{eq:model_2}:
\begin{equation}
\widehat{\mathbf{T}}(\mathbf{X}, \Delta) = a_{\mathrm{T}} \mathrm{ANN}_{\theta}(\mathbf{X} / a_{\mathrm{X}}, \Delta / a_{\Delta}), \label{eq:model_1}
\end{equation}
where $\mathrm{ANN}_{\theta}$ is the neural network with trainable parameters $\theta$. Coefficients $a_{\mathrm{T}}=$$10^{-2}$m$^2$s$^{-2}$ and $a_{\mathrm{X}}=$ $10^{-6}$s$^{-1}$ approximate the standard deviations of eddy fluxes and velocity gradients in our dataset, and $a_{\Delta}=50 \mathrm{km}$.
Using fixed normalization coefficients in parameterizations similar to Eq. \eqref{eq:model_1} is a common practice \cite{srinivasan2024turbulence}. Below, we contrast this approach to a normalization that follows solely from dimensional analysis presented in Section \ref{sec:general_algorithm}.


A combined set of input and output features ($\mathbf{T}$, $\mathbf{X}$, $\Delta$) is used to construct non-dimensional input ($\mathbf{X} / ||\mathbf{X}||_2$) and output ($\mathbf{T} / (\Delta^2 ||\mathbf{X}||_2^2)$) features (step 2 in Section \ref{sec:general_algorithm}), where $||\mathbf{X}||_2= \sqrt{\sum_i X_i^2}$. This normalization of features is local, that is computed separately for each grid point. By designing the ANN to operate on non-dimensional variables, we propose a parameterization with the local dimensional scaling \cite{reissmann2021application, prakash2022invariant}:
\begin{equation}
    \widehat{\mathbf{T}}(\mathbf{X}, \Delta) = \Delta^2 ||\mathbf{X}||_2^2 \mathrm{ANN}_{\theta} (\mathbf{X} / ||\mathbf{X}||_2). \label{eq:model_3}
\end{equation}
According to the \citeA{buckingham1914physically}'s Pi theorem, there is freedom in constructing non-dimensional variables. We opt to use the non-dimensional vector $\mathbf{X} / ||\mathbf{X}||_2$ to constrain the range of its components between $-1$ and $1$, thereby reducing the distribution shift in ANN inputs. 

Following step 3 in Section \ref{sec:general_algorithm}, we show that the model form (Eq. \eqref{eq:model_3}) admits ZB20, Smagorinsky, biharmonic Smagorinsky, and \citeA{leith1996stochastic} parameterizations as special cases, with well-behaved functional representations (see Text S1 in SI for details). Furthermore, Eq. \eqref{eq:model_3} guarantees that the predicted fluxes vanish ($\widehat{\mathbf{T}}\to\mathbf{0}$) as velocity gradients diminish ($\mathbf{X}\to\mathbf{0}$), similarly to known parameterizations, see Text S2 in SI. We experimentally verified that the spatial variability of the normalization factor ($||\mathbf{X}||_2$) in Eq. \eqref{eq:model_3} does not amplify the parameterization errors ($\mathbf{S}-\nabla \cdot \widehat{\mathbf{T}}$) compared to Eq. \eqref{eq:model_1}.

\section{Experimental setups}

\subsection{Training dataset}
The training dataset is created using the climate model CM2.6 \cite{griffies2015impacts}, which has a nominal ocean resolution of $0.1^\circ$.
Velocity gradients (Eq. \eqref{eq:ZB_features}), used as input features, and subfilter forcing ($\mathbf{S}$, Eq. \eqref{eq:subfilter_forcing}), used as output, are diagnosed using horizontal filtering followed by coarse-graining, which avoids the inclusion of discretization errors \cite{guillaumin2021stochastic,christensen2022parametrization,agdestein2025discretize}.
The filtering is applied by sliding a Gaussian kernel with a filter width three times the width of the target coarse grid box, using \citeA{grooms2021diffusion,loose2022gcm}. Subsequent coarse-graining is done by averaging over the fine grid boxes contained within each coarse grid box. 
The filtering and coarse-graining are done for 4 coarse resolutions and 10 depth levels (Figure \ref{fig:offline}(d) and Table S1 in SI). 

\subsection{ANN architecture}
In the offline analysis of ANN parameterizations (Eqs. \eqref{eq:model_1}, \eqref{eq:model_3}), we use a neural network with two hidden layers, 32 neurons each, which was found to be sufficiently large to effectively learn from the input features. See Text S3 in SI for details. 

\subsection{Online implementation} \label{sec:online_implementation}
We implement the proposed ANN mesoscale eddy parameterization (Eq. \eqref{eq:model_3}) in two considerably different configurations of the GFDL MOM6 ocean model \cite{adcroft2019gfdl} at eddy-permitting ($1/4^\circ$) resolution. To ensure that ANN inference remains computationally efficient, we retrain a smaller network with only one hidden layer and 20 neurons, which keeps the ANN inference time below 10\% of the ocean model runtime (Text S3 in SI for details). While our goal was to implement the ANN parameterization without further modifications, minor adjustments were necessary for numerical stability, see Text S4 in the SI.

The idealized ocean configuration, NeverWorld2 (NW2, \citeA{marques2022neverworld2}), includes 15 stacked shallow water layers, featuring a single basin ocean with a reentrant channel. The circulation is driven by a steady wind forcing, giving rise to a circumpolar current and gyres. Coarse simulations are initialized from rest, and run for $30000$ days, similar to \citeA{marques2022neverworld2} and \citeA{perezhogin2024stable}.

The second configuration, OM4 \cite{adcroft2019gfdl}, is a coupled ocean-sea-ice model forced at the air-sea interface by prescribing the atmosphere state according to the CORE-II interannual forcing (IAF) protocol \cite{large2009global}. The simulations span 60 years (1948-2007) and were initialized with a state of the Control model after 270 years of spin-up.

The biharmonic Smagorinsky scheme for gridscale dissipation is used with viscosity coefficient $\nu_4 = 0.06 \sqrt{\overline{\sigma}_S^2+\overline{\sigma}_T^2} \Delta^4$   \cite{adcroft2019gfdl}, applied in control and parameterized (mixed modeling, \citeA{meneveau2000scale}) simulations.

\section{Results}

\subsection{Offline generalization}
Our primary goal is to demonstrate that the local dimensional scaling promotes the generalization of the eddy parameterization to unseen grid resolutions and depths. Limited generalization in such scenarios has been reported for previous machine-learning models of mesoscale eddies \cite{zhang2023implementation, gultekin2024analysis,ross2022benchmarking} and traditional physics-based parameterizations \cite{yankovsky2024}. 

We compare two ANNs: one incorporating local dimensional scaling (Eq. \eqref{eq:model_3}) and a baseline ANN with fixed normalization coefficients (Eq. \eqref{eq:model_1}). To explore generalization, we let the ANNs learn based solely on data from one combination of depth ($5$m) and coarse grid resolution ($0.9^\circ$) during training. The local grid spacing varies according to the tripolar grid used in the ocean component of the CM2.6 climate model. In particular, at the nominal resolution of $0.9^\circ$, the coarse grid spacing $\Delta=\sqrt{\Delta x \Delta y}$ is in a range from 50km to 100km for non-polar latitudes ($60\mathrm{S}^\circ-60\mathrm{N}^\circ$). Spatially varying grid spacing provides essential information for effective learning by the baseline ANN. The offline evaluation of ANNs on held-out data similar to that used for training is shown in Figure \ref{fig:offline}(b). Both ANNs exhibit high and equal pattern correlation ($0.90$) and $\mathrm{R}^2$ ($0.81$) in the prediction of the norm of subfilter forcing. 

We now consider generalization to a different grid resolution, which is finer ($0.4^\circ$) compared to that used for training ($0.9^\circ$), see Figure \ref{fig:offline}(c). The range of grid spacings in this case is beyond the range seen by a baseline ANN during training, resulting in a distribution shift between the testing and training data.  The baseline ANN parameterization (Eq. \eqref{eq:model_1}) predicts the norm of subfilter forcing at a new grid resolution with a reasonably high pattern correlation ($0.85$). However, the magnitude of the prediction is too large, resulting in a low  $\mathrm{R}^2$ ($-2.71$). 
Instead, the ANN with dimensional scaling (Eq. \eqref{eq:model_3}) offers improved generalization capability. The proposed ANN naturally accounts for the reduction of the grid spacing and reduces the magnitude of the prediction, resulting in high pattern correlation ($0.94$) and $\mathrm{R}^2$ (0.87) metrics (Figure \ref{fig:offline}(c)). 

The generalization to both finer and coarser grids, and different depths, is summarized in Figure \ref{fig:offline}(d). At coarser grid spacings ($1.2^\circ$, $1.5^\circ$) compared to that used for training ($0.9^\circ$), the local dimensional scaling again helps to achieve higher $\mathrm{R}^2$ by increasing the magnitude of the prediction. 

In the deep layers, the subfilter forcing and velocity gradients are approximately one order of magnitude smaller than near the surface. Thus, a baseline ANN, trained on much larger values near the surface (such as here, at depth $5$m), can lead to inaccurate predictions at depth (Figure S1 in SI). However, the local dimensional scaling effectively rescales the input features, thereby improving generalization to deep, unseen layers, as summarized in Figure \ref{fig:offline}(d). 

The major reason why baseline ANN has a poor skill at unseen resolutions and depths is the lack of training data. We verified that the generalization of the baseline ANN to various resolutions and depths can be restored if these resolutions and depths are included in the training dataset. Incorporating local dimensional scaling as done in this work, therefore, requires less training data and improves out-of-distribution generalization.

\subsection{Online evaluation in the MOM6 ocean model}
\begin{figure}[h!]
\centering{\includegraphics[width=0.8\textwidth]{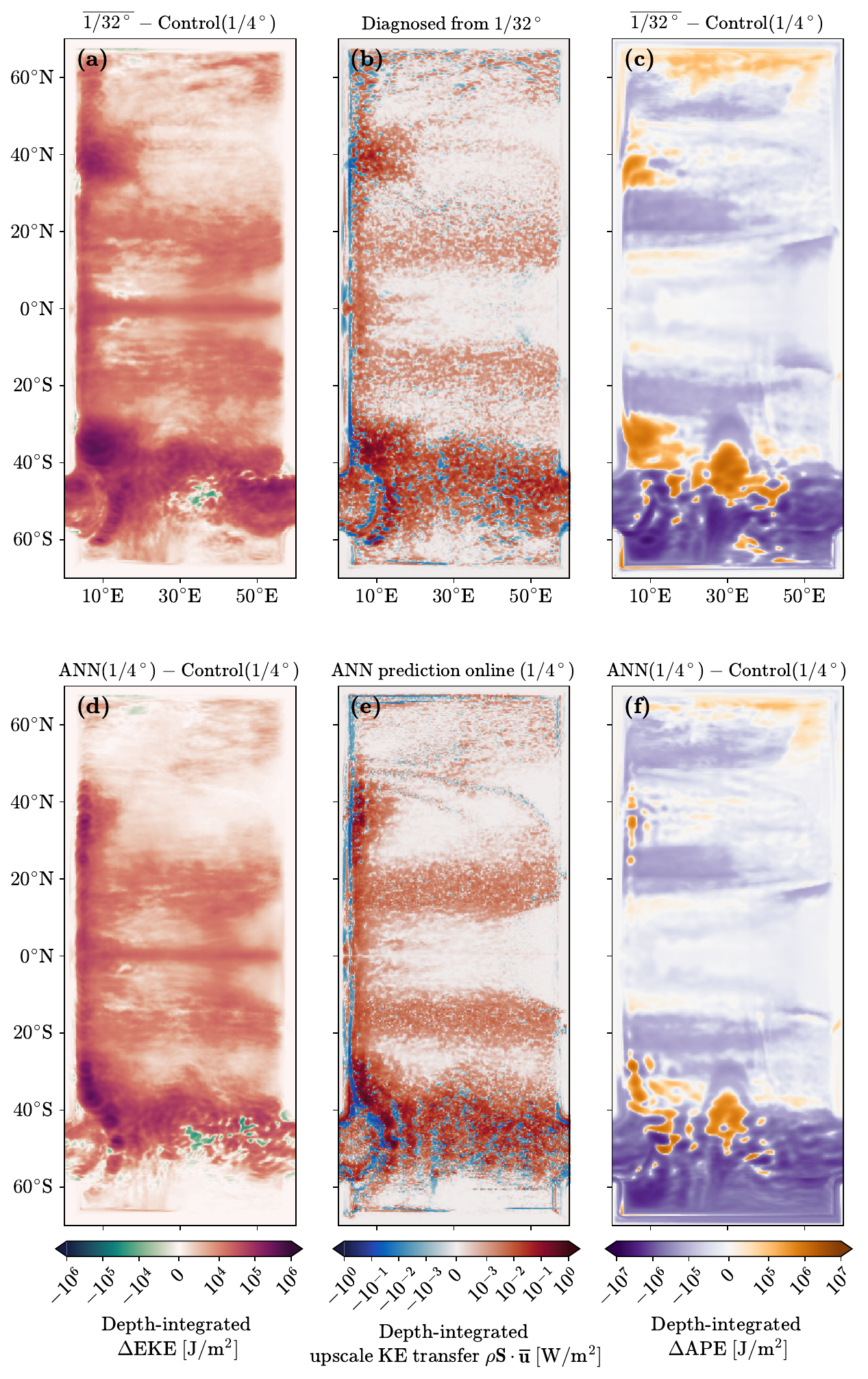}}
\caption{
Effect of increasing resolution in idealized configuration NeverWorld2 \cite{marques2022neverworld2} in the upper row, where $\overline{1/32^\circ}$ represents filtered and coarse-grained high-resolution simulation.  Impact of the ANN parameterization with local dimensional scaling (Eq. \eqref{eq:model_3}) online at resolution $1/4^\circ$ in the lower row.
We consider three depth-integrated metrics: difference (denoted as  $\Delta$) in Eddy Kinetic Energy (EKE) (left column); diagnosed and predicted online upscale kinetic energy transfer (center column) (positive values represent backscatter);  difference in Available Potential Energy (APE) (right column). All metrics are averaged over 160 snapshots corresponding to the last 800 days of the simulations.
}
\label{fig:online-1}
\end{figure} 

We use all available depths and grid resolutions shown in Figure \ref{fig:offline}(d) for training to make the implemented ANN parameterization (Eq. \eqref{eq:model_3}) less tied to any specific resolution or depth. The retrained, smaller ANN (see Section \ref{sec:online_implementation}) exhibits a slightly lower offline skill ($R^2$) than the version described earlier, on average, by 0.1 (Figure S5 in SI).

\subsubsection{Idealized configuration NeverWorld2} \label{sec:nw2}

We first consider an idealized adiabatic ocean configuration NW2, which generates various circulation patterns similar to the global ocean but allows us to isolate the effect of mesoscale eddies. Our goal is to show that the impact of the ANN parameterization on the flow is similar to that of increasing the horizontal resolution. 

Mesoscale eddies extract available potential energy, APE$=\frac{\rho}{2} \sum_k g_k' ( \eta_k^2 - (\eta^{\mathrm{ref}}_k)^2 )$, from the mean flow, which is then converted into the eddy kinetic energy, EKE$=\frac{\rho}{2}(\overline{|\mathbf{u}^2|}^t-|\overline{\mathbf{u}}^t|^2)$ \cite{salmon1980baroclinic}. Here, $\overline{(\cdot)}^t$ is the temporal averaging, $\rho$ is the density, $g'_k$ is the reduced gravity of $k$-th isopycnal interface, $\eta_k$ is the interface height and $\eta^{\mathrm{ref}}_k$ is the state of rest with flat isopycnals. At an eddy-permitting resolution ($1/4^\circ$), this energy pathway is partially unresolved \cite{jansen2014parameterizing, mana2014toward, juricke2019ocean, loose2023diagnosing}. As a result, the coarse ocean model has too low EKE and too large APE when compared to the filtered and coarse-grained high-resolution simulation, denoted as $\overline{1/32^\circ}$ (Figure \ref{fig:online-1}(a,c)). However, Figure \ref{fig:online-1}(a) suggests that the missing eddies can be nominally resolved on the coarse grid. Traditional backscatter parameterizations are designed to directly reduce this EKE bias by energizing the resolved eddies, resulting in additional extraction of APE \cite{jansen2014parameterizing, yankovsky2024}.

Eddy backscatter is diagnosed when the kinetic energy transfer produced by the subfilter forcing (Eq. \eqref{eq:subfilter_forcing}) is predominantly positive (upscale), as shown in Figure \ref{fig:online-1}(b).
We verified that our ANN parameterization accurately predicts the eddy backscatter offline (Figure S2 in SI), suggesting reasonable generalization capabilities of our approach. In Figure \ref{fig:online-1}(e), we show a more challenging task -- the prediction of the eddy backscatter online once the ANN is coupled to the coarse ocean model. The online prediction of eddy backscatter grossly resembles the diagnosed data shown in Figure \ref{fig:online-1}(b), although there are slight differences caused by the difference in distributions of input features.

The kinetic energy injection from the ANN parameterization leads to an increase in EKE that aligns with the high-resolution data, in particular in the ACC (Antarctic circumpolar current) region ($40^\circ \mathrm{S}-60^\circ\mathrm{S}$), near the western boundaries, and in the subtropics ($20^\circ \mathrm{S}-20^\circ\mathrm{N}$), see Figure \ref{fig:online-1}(d). 
However, the EKE increase in western boundary current extension ($40^\circ\mathrm{N}$, $10^\circ\mathrm{E}-20^\circ\mathrm{E}$) and subpolar gyre ($50^\circ\mathrm{N}-70^\circ\mathrm{N}$) is smaller than expected from the high-resolution simulation. The spatial pattern of APE reduction in the parameterized simulation is close to that produced by increasing horizontal resolution (Figure \ref{fig:online-1}(f)). The APE is predominantly reduced in the Southern Ocean and ACC regions ($40^\circ \mathrm{S}-70^\circ\mathrm{S}$), followed by APE reduction in gyres ($20^\circ \mathrm{N}-60^\circ\mathrm{N}$, $20^\circ \mathrm{S}-40^\circ\mathrm{S}$). Local patches of  APE increase in the higher resolution model (Figure \ref{fig:online-1}(c)) correspond to enhanced horizontal recirculation and are reproduced by the ANN parameterization, but less accurately compared to the diagnosed APE reduction.

Consequences of APE reduction include flattening of isopycnals and improving the structure of isopycnal interfaces across multiple cross-sections (Figure S3 in SI), along with a weakening of ACC transport through the Drake Passage (Table S3 in SI).

\begin{figure}[h!]
\centering{\includegraphics[width=0.9\textwidth]{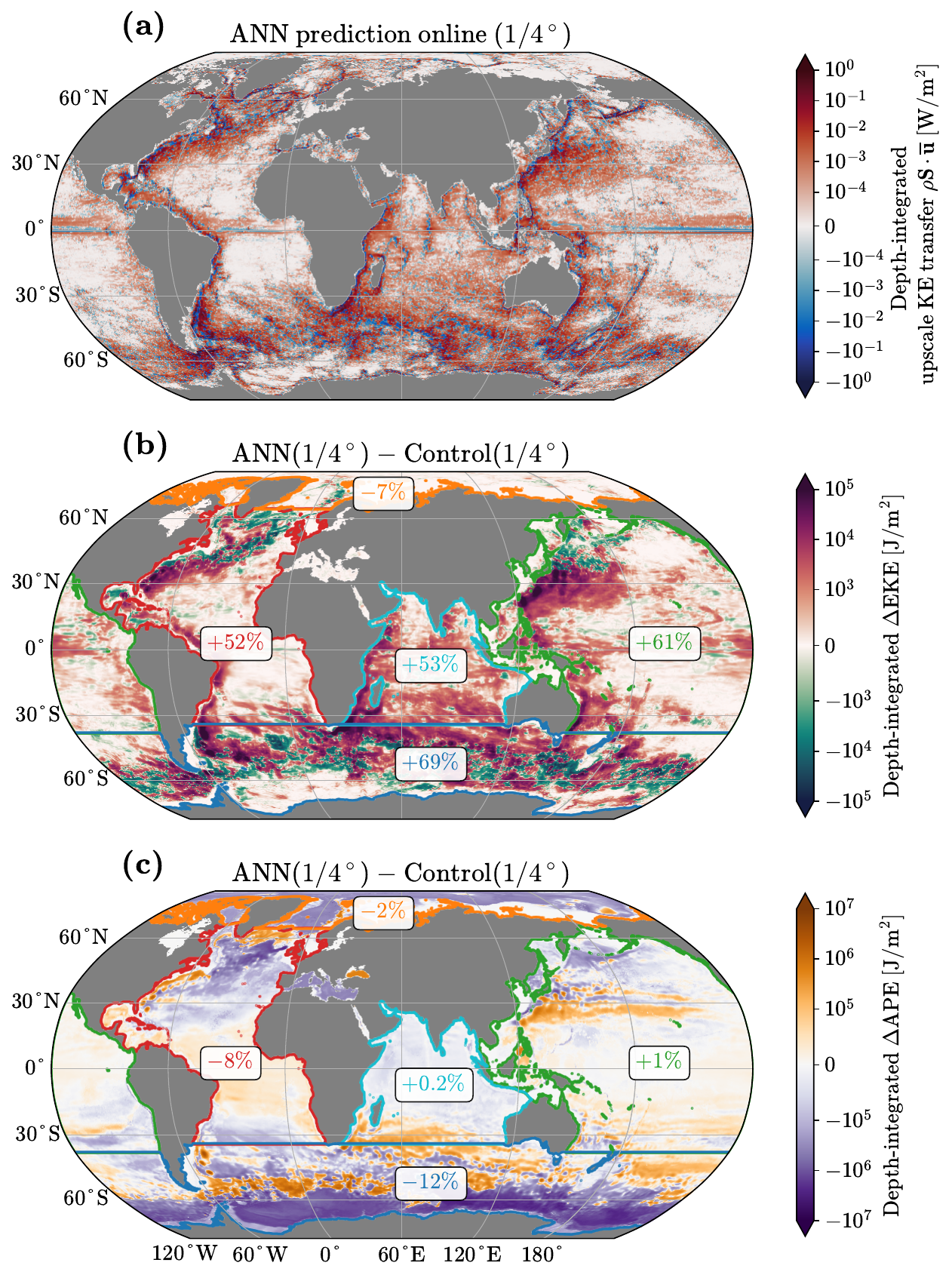}}
\caption{Online evaluation of the ANN parameterization in the global ocean-ice model OM4 \cite{adcroft2019gfdl} at eddy-permitting resolution ($1/4^\circ$). The following depth-integrated diagnostics are averaged over one year (2003): (a) upscale kinetic energy transfer predicted by the ANN parameterization online, (b) difference in Eddy kinetic energy (EKE), (c) difference in Available potential energy (APE). The integrated percentage change in EKE and APE relative to the control simulation is shown for five ocean basins.
}
\label{fig:online-2}
\end{figure}

\subsubsection{Global ocean-sea-ice model OM4}
We next evaluate the ANN parameterization in the global ocean model OM4. Unlike in the idealized configuration, the interaction of many physical processes in driving the circulation in the global ocean model impede our ability to directly isolate the effect of mesoscale eddies \cite{ferrari2009ocean, levy2010modifications}. Building on the dynamical expectations established in the idealized NW2 configuration, our goal is to assess whether the global ocean model exhibits similar response patterns to the eddy parameterization.

The prediction of the kinetic energy injection by the ANN parameterization online is shown in Figure \ref{fig:online-2}(a). Similarly to the idealized configuration, the kinetic energy is injected in the subtropical gyres, 
near the western boundaries, and occasionally in the ACC region. The energy injection is accompanied by an increase of the EKE in the same locations, see Figure \ref{fig:online-2}(b). However, compared to the pattern found in an idealized configuration, the EKE decrease appears more frequently: along topographic features in the subpolar gyres of the North Atlantic and North Pacific oceans, and occasionally in the ACC region. The decrease of EKE in these regions is due to a shift or weakening of the mean currents, potentially as a result of the removal of kinetic energy by the ANN parameterization along the lateral boundaries, changes in deep water formation, and/or changes in global overturning circulation. The complexity of the model prohibits us from identifying a single mechanism. 

Similarly to the idealized configuration, APE is primarily reduced in the Southern Ocean ($-12 \%$), with minor APE reductions observed in the Subpolar Gyres of the North Atlantic and North Pacific oceans (Figure \ref{fig:online-2}(c)). APE is additionally reduced in the Arctic Ocean despite the lack of increased eddy activity in this region. However, its relative change is moderately small ($-2 \%$).

We assessed whether the offline performance of the ANN parameterization correlates with the online results (Figure S5 in SI). We found that using spatially non-local features on a $3 \times 3$ stencil, as in this study, is important for achieving higher offline skill in CM2.6 and improved energetics in OM4 compared to a pointwise ANN parameterization and ZB20 closure. However, increasing the number of neurons, which also contributes to the offline skill, has a smaller impact on the energetics. Note that the traditional anti-viscosity parameterization is capable of improving energetics as well, while its offline skill is close to zero (Figure S5 in SI, see also \citeA{ross2022benchmarking}).

We note that the evaluation presented in this section is qualitative and can be strengthened by comparing the parameterized global ocean model to filtered and coarse-grained higher-resolution simulations. Such evaluation can be performed in future studies by implementing the proposed parameterization in the hierarchy of GFDL climate models, CM4X, which differ in the horizontal resolution of the ocean component \cite{griffies2024cm4xpart1}.

\subsubsection{Comparison to an anti-viscosity parameterization}
\begin{figure}[h!]
\centering{\includegraphics[width=0.9\textwidth]{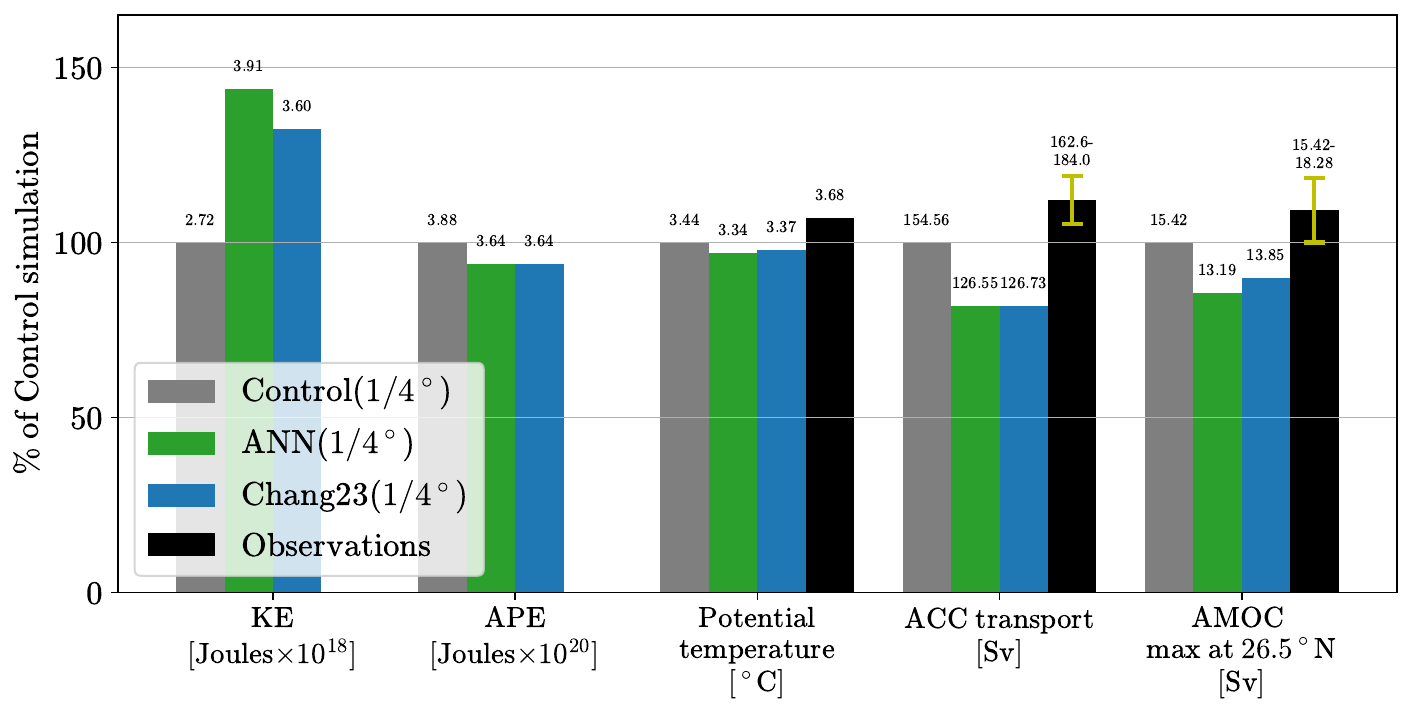}}
\caption{Comparison of the ANN parameterization to the negative viscosity backscatter parameterization \cite{chang2023remote} in the global ocean-ice model OM4. Kinetic energy (KE) and Available potential energy (APE) are integrated globally. Potential temperature is averaged globally. ACC transport is computed at the Drake Passage section at $70^\circ$W, and AMOC is computed as the maximum over depth streamfunction at $26.5^\circ$N. Model output is averaged over the years 1981-2007. Observational data for potential temperature is given by World Ocean Atlas 2018 (WOA18, \citeA{locarnini2018world}), for ACC transport with error bar is given by cDrake \cite{donohue2016mean}, and for AMOC is given by RAPID \cite{cunningham2007temporal}  averaged over 2004-2021 years with error bar showing interannual standard deviation.
}
\label{fig:online_metrics}
\end{figure}

We confront our ANN parameterization to a traditional anti-viscosity parameterization representing mesoscale eddy effects \cite{jansen2015energy} and already tested in OM4 by \citeA{chang2023remote}. Repeating their analysis, we found that both ANN and anti-viscosity parameterizations reduce the regional biases in the Gulf Stream region, see Figure S4 in SI for sea surface temperature and salinity biases. The response in other global ocean circulation metrics is remarkably similar for both parameterizations as well (Figure \ref{fig:online_metrics}). Specifically, both parameterizations increase the globally integrated kinetic energy by roughly the same percentage and reduce the APE by nearly the same percentage. The restratification effect of mesoscale eddies leads to the reduction of the globally-averaged potential temperature \cite{griffies2015impacts, adcroft2019gfdl}. As previously discussed, the transport through the Drake Passage is reduced in both parametrized simulations, see also \citeA{grooms2024stochastic}. Unlike in \citeA{chang2023remote}, both parameterizations weaken the Atlantic meridional circulation (AMOC). This suggests that the AMOC response depends on the ocean model state, perhaps to a greater extent than the details of mesoscale eddy parameterizations. The response in some global metrics (ACC, AMOC, globally-averaged potential temperature) does not appear to project onto the existing ocean model biases. 
That is, both parameterized ocean simulations are less consistent with the observational data than the control simulation (Figure \ref{fig:online_metrics}). 
We note that our goal was to improve the representation of mesoscale eddy processes. Bias reduction is not guaranteed due to compensating model errors from other parameterizations and remains an important direction for future work. A full recalibration of the ocean model may be necessary, particularly for physical processes competing with mesoscale eddies in determining average potential temperature and the strength of the ACC and AMOC.


\section{Discussion}

We address the generalization issue of ANN parameterizations of mesoscale eddies by embedding physics constraints into the inputs, outputs, and parametrization itself. The \citeA{buckingham1914physically}'s Pi-theorem and dimensional analysis are invoked to obtain local normalization coefficients. The ANN parameterization with local dimensional scaling significantly outperforms the ANN with fixed normalization coefficients offline, demonstrating superior generalization to unseen grid resolutions and depths in the global ocean data CM2.6. A general algorithm for constructing dimensional scaling, which can be applied to other neural-network parameterizations, is presented.

The proposed ANN parameterization with dimensional scaling is successfully tested online in the GFDL MOM6 ocean model. It accurately predicts upscale kinetic energy transfer, despite many challenges presented by online implementation. The parameterization improves the energy pathways by energizing the resolved eddies and reducing APE, consistent with the expected restratification effects of mesoscale eddies. These improvements hold across idealized (NW2) and global ocean (OM4) configurations, with the most pronounced APE reduction occurring in the Southern Ocean. The ANN achieves comparable online performance to an existing backscatter parameterization \cite{jansen2015energy, chang2023remote} in OM4, and does not require significant retuning between idealized and global setups.

We demonstrate the improved or similar performance of the ANN parameterization in NW2 compared to existing backscatter schemes \cite{yankovsky2024, perezhogin2024stable} across different resolutions ($1/3^\circ-1/6^\circ$, see Table S2 in SI). At $1/2^\circ$, however, the ANN offers no clear improvement compared to the control simulation, likely due to less resolved eddies and stronger viscosity. At coarser resolutions ($\sim1^\circ$),  the subfilter momentum fluxes vanish as the Rossby radius is unresolved (Figure S6 in SI). At such coarse resolutions, combining the ANN with bulk parameterizations or online learning approaches may help \cite{maddison2024online, shankar2025differentiable}, along with parameterizations explicitly extracting APE \cite{bachman2019gm, jansen2019toward, grooms2024stochastic, perezhogin2025large, balwada2025design}.


Additional work is needed to enhance data-driven parameterizations beyond the performance of traditional parameterizations in realistic global configurations. Both parameterization approaches exhibit substantial departures from observations and contribute comparably to persistent model biases. This highlights the potential for improving parameterization schemes, evaluation metrics, and model calibration in ocean modeling. Looking ahead, the generalization issue addressed in this study has immediate implications for climate models, where parameterizations must remain reliable under changing conditions.

\section*{Open Research Section}
The training algorithm, plots, ANN weights, implemented parameterization and MOM6 setups are available at \citeA{perezhogin_2025_software}. The training dataset, offline skill, and simulation data are available at \citeA{perezhogin_2025_dataset}. For high-resolution NW2 simulation data, see \citeA{NW2data}. Observational products can be found: WOA18 \cite{WOA18data} and RAPID \cite{RAPIDdata}.

\acknowledgments
This project is supported by Schmidt Sciences, LLC. 
LZ was also partially funded through NOAA NA19OAR4310364-T1-01. 
This research was also supported in part through the NYU IT High Performance
Computing resources, services, and staff expertise. This research used resources of the National Energy Research Scientific Computing Center, a DOE Office of Science User Facility supported by the Office of Science of the U.S. Department of Energy under Contract No. DE-AC02-05CH11231 using NERSC award BER-ERCAP0032655. The authors would like to thank the M$^2$LInES team, in particular, Dhruv Balwada, Alex Connolly, and Nora Loose, and also Wenda Zhang and Aviral Prakash for helpful comments and discussion. We also thank the anonymous reviewers for their helpful suggestions, which have helped improve the manuscript. 
%
%


\bibliography{agusample}

\begin{thebibliography}{}

\bibitem [\protect \citeauthoryear {%
Adcroft%
\ \protect \BOthers {.}}{%
Adcroft%
\ \protect \BOthers {.}}{%
{\protect \APACyear {2019}}%
}]{%
adcroft2019gfdl}
\APACinsertmetastar {%
adcroft2019gfdl}%
\begin{APACrefauthors}%
Adcroft, A.%
, Anderson, W.%
, Balaji, V.%
, Blanton, C.%
, Bushuk, M.%
, Dufour, C\BPBI O.%
\BDBL {}others%
\end{APACrefauthors}%
\unskip\
\newblock
\APACrefYearMonthDay{2019}{}{}.
\newblock
{\BBOQ}\APACrefatitle {{The GFDL global ocean and sea ice model OM4.0: Model description and simulation features}} {{The GFDL global ocean and sea ice model OM4.0: Model description and simulation features}}.{\BBCQ}
\newblock
\APACjournalVolNumPages{Journal of Advances in Modeling Earth Systems}{11}{10}{3167--3211}.
\newblock
\begin{APACrefDOI} \doi{https://doi.org/10.1029/2019MS001726} \end{APACrefDOI}
\PrintBackRefs{\CurrentBib}

\bibitem [\protect \citeauthoryear {%
Agdestein%
\ \BBA {} Sanderse%
}{%
Agdestein%
\ \BBA {} Sanderse%
}{%
{\protect \APACyear {2025}}%
}]{%
agdestein2025discretize}
\APACinsertmetastar {%
agdestein2025discretize}%
\begin{APACrefauthors}%
Agdestein, S\BPBI D.%
\BCBT {}\ \BBA {} Sanderse, B.%
\end{APACrefauthors}%
\unskip\
\newblock
\APACrefYearMonthDay{2025}{}{}.
\newblock
{\BBOQ}\APACrefatitle {Discretize first, filter next: Learning divergence-consistent closure models for large-eddy simulation} {Discretize first, filter next: Learning divergence-consistent closure models for large-eddy simulation}.{\BBCQ}
\newblock
\APACjournalVolNumPages{Journal of Computational Physics}{522}{}{113577}.
\newblock
\begin{APACrefDOI} \doi{https://doi.org/10.1016/j.jcp.2024.113577} \end{APACrefDOI}
\PrintBackRefs{\CurrentBib}

\bibitem [\protect \citeauthoryear {%
Bachman%
}{%
Bachman%
}{%
{\protect \APACyear {2019}}%
}]{%
bachman2019gm}
\APACinsertmetastar {%
bachman2019gm}%
\begin{APACrefauthors}%
Bachman, S\BPBI D.%
\end{APACrefauthors}%
\unskip\
\newblock
\APACrefYearMonthDay{2019}{}{}.
\newblock
{\BBOQ}\APACrefatitle {{The GM + E closure: A framework for coupling backscatter with the Gent and McWilliams parameterization}} {{The GM + E closure: A framework for coupling backscatter with the Gent and McWilliams parameterization}}.{\BBCQ}
\newblock
\APACjournalVolNumPages{Ocean Modelling}{136}{}{85--106}.
\newblock
\begin{APACrefDOI} \doi{https://doi.org/10.1016/j.ocemod.2019.02.006} \end{APACrefDOI}
\PrintBackRefs{\CurrentBib}

\bibitem [\protect \citeauthoryear {%
Bachman%
, Fox-Kemper%
\BCBL {}\ \BBA {} Pearson%
}{%
Bachman%
\ \protect \BOthers {.}}{%
{\protect \APACyear {2017}}%
}]{%
bachman2017scale}
\APACinsertmetastar {%
bachman2017scale}%
\begin{APACrefauthors}%
Bachman, S\BPBI D.%
, Fox-Kemper, B.%
\BCBL {}\ \BBA {} Pearson, B.%
\end{APACrefauthors}%
\unskip\
\newblock
\APACrefYearMonthDay{2017}{}{}.
\newblock
{\BBOQ}\APACrefatitle {A scale-aware subgrid model for quasi-geostrophic turbulence} {A scale-aware subgrid model for quasi-geostrophic turbulence}.{\BBCQ}
\newblock
\APACjournalVolNumPages{Journal of Geophysical Research: Oceans}{122}{2}{1529--1554}.
\newblock
\begin{APACrefDOI} \doi{https://doi.org/10.1002/2016JC012265} \end{APACrefDOI}
\PrintBackRefs{\CurrentBib}

\bibitem [\protect \citeauthoryear {%
Balwada%
, Perezhogin%
, Adcroft%
\BCBL {}\ \BBA {} Zanna%
}{%
Balwada%
\ \protect \BOthers {.}}{%
{\protect \APACyear {2025}}%
}]{%
balwada2025design}
\APACinsertmetastar {%
balwada2025design}%
\begin{APACrefauthors}%
Balwada, D.%
, Perezhogin, P.%
, Adcroft, A.%
\BCBL {}\ \BBA {} Zanna, L.%
\end{APACrefauthors}%
\unskip\
\newblock
\APACrefYearMonthDay{2025}{}{}.
\newblock
{\BBOQ}\APACrefatitle {Design and implementation of a data-driven parameterization for mesoscale thickness fluxes} {Design and implementation of a data-driven parameterization for mesoscale thickness fluxes}.{\BBCQ}
\newblock
\APACjournalVolNumPages{Authorea Preprints}{}{}{}.
\PrintBackRefs{\CurrentBib}

\bibitem [\protect \citeauthoryear {%
Barenblatt%
}{%
Barenblatt%
}{%
{\protect \APACyear {1996}}%
}]{%
barenblatt1996scaling}
\APACinsertmetastar {%
barenblatt1996scaling}%
\begin{APACrefauthors}%
Barenblatt, G\BPBI I.%
\end{APACrefauthors}%
\unskip\
\newblock
\APACrefYear{1996}.
\newblock
\APACrefbtitle {Scaling, self-similarity, and intermediate asymptotics: dimensional analysis and intermediate asymptotics} {Scaling, self-similarity, and intermediate asymptotics: dimensional analysis and intermediate asymptotics}\ (\BNUM~14).
\newblock
\APACaddressPublisher{}{Cambridge University Press}.
\PrintBackRefs{\CurrentBib}

\bibitem [\protect \citeauthoryear {%
Beucler%
\ \protect \BOthers {.}}{%
Beucler%
\ \protect \BOthers {.}}{%
{\protect \APACyear {2024}}%
}]{%
beucler2024climate}
\APACinsertmetastar {%
beucler2024climate}%
\begin{APACrefauthors}%
Beucler, T.%
, Gentine, P.%
, Yuval, J.%
, Gupta, A.%
, Peng, L.%
, Lin, J.%
\BDBL {}others%
\end{APACrefauthors}%
\unskip\
\newblock
\APACrefYearMonthDay{2024}{}{}.
\newblock
{\BBOQ}\APACrefatitle {Climate-invariant machine learning} {Climate-invariant machine learning}.{\BBCQ}
\newblock
\APACjournalVolNumPages{Science Advances}{10}{6}{eadj7250}.
\newblock
\begin{APACrefDOI} \doi{https://doi.org/10.1126/sciadv.adj7250} \end{APACrefDOI}
\PrintBackRefs{\CurrentBib}

\bibitem [\protect \citeauthoryear {%
Bishop%
\ \BBA {} Nasrabadi%
}{%
Bishop%
\ \BBA {} Nasrabadi%
}{%
{\protect \APACyear {2006}}%
}]{%
bishop2006pattern}
\APACinsertmetastar {%
bishop2006pattern}%
\begin{APACrefauthors}%
Bishop, C\BPBI M.%
\BCBT {}\ \BBA {} Nasrabadi, N\BPBI M.%
\end{APACrefauthors}%
\unskip\
\newblock
\APACrefYear{2006}.
\newblock
\APACrefbtitle {Pattern recognition and machine learning} {Pattern recognition and machine learning}\ (\BVOL~4)\ (\BNUM~4).
\newblock
\APACaddressPublisher{}{Springer}.
\PrintBackRefs{\CurrentBib}

\bibitem [\protect \citeauthoryear {%
Bolton%
\ \BBA {} Zanna%
}{%
Bolton%
\ \BBA {} Zanna%
}{%
{\protect \APACyear {2019}}%
}]{%
Bolton2019}
\APACinsertmetastar {%
Bolton2019}%
\begin{APACrefauthors}%
Bolton, T.%
\BCBT {}\ \BBA {} Zanna, L.%
\end{APACrefauthors}%
\unskip\
\newblock
\APACrefYearMonthDay{2019}{}{}.
\newblock
{\BBOQ}\APACrefatitle {{Applications of Deep Learning to Ocean Data Inference and Subgrid Parameterization}} {{Applications of Deep Learning to Ocean Data Inference and Subgrid Parameterization}}.{\BBCQ}
\newblock
\APACjournalVolNumPages{Journal of Advances in Modeling Earth Systems}{11}{1}{376--399}.
\newblock
\begin{APACrefDOI} \doi{https://doi.org/10.1029/2018MS001472} \end{APACrefDOI}
\PrintBackRefs{\CurrentBib}

\bibitem [\protect \citeauthoryear {%
Bridgman%
}{%
Bridgman%
}{%
{\protect \APACyear {1922}}%
}]{%
bridgman1922dimensional}
\APACinsertmetastar {%
bridgman1922dimensional}%
\begin{APACrefauthors}%
Bridgman, P\BPBI W.%
\end{APACrefauthors}%
\unskip\
\newblock
\APACrefYear{1922}.
\newblock
\APACrefbtitle {Dimensional analysis} {Dimensional analysis}.
\newblock
\APACaddressPublisher{}{Yale university press}.
\PrintBackRefs{\CurrentBib}

\bibitem [\protect \citeauthoryear {%
Buckingham%
}{%
Buckingham%
}{%
{\protect \APACyear {1914}}%
}]{%
buckingham1914physically}
\APACinsertmetastar {%
buckingham1914physically}%
\begin{APACrefauthors}%
Buckingham, E.%
\end{APACrefauthors}%
\unskip\
\newblock
\APACrefYearMonthDay{1914}{}{}.
\newblock
{\BBOQ}\APACrefatitle {On physically similar systems; illustrations of the use of dimensional equations} {On physically similar systems; illustrations of the use of dimensional equations}.{\BBCQ}
\newblock
\APACjournalVolNumPages{Physical review}{4}{4}{345}.
\newblock
\begin{APACrefDOI} \doi{https://doi.org/10.1103/PhysRev.4.345} \end{APACrefDOI}
\PrintBackRefs{\CurrentBib}

\bibitem [\protect \citeauthoryear {%
Chang%
, Adcroft%
, Zanna%
, Hallberg%
\BCBL {}\ \BBA {} Griffies%
}{%
Chang%
\ \protect \BOthers {.}}{%
{\protect \APACyear {2023}}%
}]{%
chang2023remote}
\APACinsertmetastar {%
chang2023remote}%
\begin{APACrefauthors}%
Chang, C\BHBI Y.%
, Adcroft, A.%
, Zanna, L.%
, Hallberg, R.%
\BCBL {}\ \BBA {} Griffies, S\BPBI M.%
\end{APACrefauthors}%
\unskip\
\newblock
\APACrefYearMonthDay{2023}{}{}.
\newblock
{\BBOQ}\APACrefatitle {{Remote versus local impacts of energy backscatter on the North Atlantic SST biases in a global ocean model}} {{Remote versus local impacts of energy backscatter on the North Atlantic SST biases in a global ocean model}}.{\BBCQ}
\newblock
\APACjournalVolNumPages{{Geophysical Research Letters}}{50}{21}{e2023GL105757}.
\newblock
\begin{APACrefDOI} \doi{https://doi.org/10.1029/2023GL105757} \end{APACrefDOI}
\PrintBackRefs{\CurrentBib}

\bibitem [\protect \citeauthoryear {%
Christensen%
\ \BBA {} Zanna%
}{%
Christensen%
\ \BBA {} Zanna%
}{%
{\protect \APACyear {2022}}%
}]{%
christensen2022parametrization}
\APACinsertmetastar {%
christensen2022parametrization}%
\begin{APACrefauthors}%
Christensen, H.%
\BCBT {}\ \BBA {} Zanna, L.%
\end{APACrefauthors}%
\unskip\
\newblock
\APACrefYearMonthDay{2022}{}{}.
\newblock
{\BBOQ}\APACrefatitle {{Parametrization in Weather and Climate Models}} {{Parametrization in Weather and Climate Models}}.{\BBCQ}
\newblock
\APACaddressPublisher{}{Oxford University Press}.
\newblock
\begin{APACrefDOI} \doi{https://doi.org/10.1093/acrefore/9780190228620.013.826} \end{APACrefDOI}
\PrintBackRefs{\CurrentBib}

\bibitem [\protect \citeauthoryear {%
Christopoulos%
\ \protect \BOthers {.}}{%
Christopoulos%
\ \protect \BOthers {.}}{%
{\protect \APACyear {2024}}%
}]{%
christopoulos2024online}
\APACinsertmetastar {%
christopoulos2024online}%
\begin{APACrefauthors}%
Christopoulos, C.%
, Lopez-Gomez, I.%
, Beucler, T.%
, Cohen, Y.%
, Kawczynski, C.%
, Dunbar, O\BPBI R.%
\BCBL {}\ \BBA {} Schneider, T.%
\end{APACrefauthors}%
\unskip\
\newblock
\APACrefYearMonthDay{2024}{}{}.
\newblock
{\BBOQ}\APACrefatitle {Online learning of entrainment closures in a hybrid machine learning parameterization} {Online learning of entrainment closures in a hybrid machine learning parameterization}.{\BBCQ}
\newblock
\APACjournalVolNumPages{Journal of Advances in Modeling Earth Systems}{16}{11}{e2024MS004485}.
\newblock
\begin{APACrefDOI} \doi{https://doi.org/10.1029/2024MS004485} \end{APACrefDOI}
\PrintBackRefs{\CurrentBib}

\bibitem [\protect \citeauthoryear {%
Connolly%
\ \protect \BOthers {.}}{%
Connolly%
\ \protect \BOthers {.}}{%
{\protect \APACyear {2025}}%
}]{%
connolly2025deep}
\APACinsertmetastar {%
connolly2025deep}%
\begin{APACrefauthors}%
Connolly, A.%
, Cheng, Y.%
, Walters, R.%
, Wang, R.%
, Yu, R.%
\BCBL {}\ \BBA {} Gentine, P.%
\end{APACrefauthors}%
\unskip\
\newblock
\APACrefYearMonthDay{2025}{}{}.
\newblock
{\BBOQ}\APACrefatitle {Deep Learning Turbulence Closures Generalize Best With Physics-based Methods} {Deep learning turbulence closures generalize best with physics-based methods}.{\BBCQ}
\newblock
\APACjournalVolNumPages{Authorea Preprints}{}{}{}.
\newblock
\begin{APACrefDOI} \doi{https://doi.org/10.22541/essoar.173869578.80400701/v1} \end{APACrefDOI}
\PrintBackRefs{\CurrentBib}

\bibitem [\protect \citeauthoryear {%
Cunningham%
\ \protect \BOthers {.}}{%
Cunningham%
\ \protect \BOthers {.}}{%
{\protect \APACyear {2007}}%
}]{%
cunningham2007temporal}
\APACinsertmetastar {%
cunningham2007temporal}%
\begin{APACrefauthors}%
Cunningham, S\BPBI A.%
, Kanzow, T.%
, Rayner, D.%
, Baringer, M\BPBI O.%
, Johns, W\BPBI E.%
, Marotzke, J.%
\BDBL {}others%
\end{APACrefauthors}%
\unskip\
\newblock
\APACrefYearMonthDay{2007}{}{}.
\newblock
{\BBOQ}\APACrefatitle {{Temporal variability of the Atlantic meridional overturning circulation at 26.5 N}} {{Temporal variability of the Atlantic meridional overturning circulation at 26.5 N}}.{\BBCQ}
\newblock
\APACjournalVolNumPages{science}{317}{5840}{935--938}.
\newblock
\begin{APACrefDOI} \doi{https://doi.org/10.1126/science.1141304} \end{APACrefDOI}
\PrintBackRefs{\CurrentBib}

\bibitem [\protect \citeauthoryear {%
Donohue%
, Tracey%
, Watts%
, Chidichimo%
\BCBL {}\ \BBA {} Chereskin%
}{%
Donohue%
\ \protect \BOthers {.}}{%
{\protect \APACyear {2016}}%
}]{%
donohue2016mean}
\APACinsertmetastar {%
donohue2016mean}%
\begin{APACrefauthors}%
Donohue, K\BPBI A.%
, Tracey, K.%
, Watts, D\BPBI R.%
, Chidichimo, M\BPBI P.%
\BCBL {}\ \BBA {} Chereskin, T.%
\end{APACrefauthors}%
\unskip\
\newblock
\APACrefYearMonthDay{2016}{}{}.
\newblock
{\BBOQ}\APACrefatitle {{Mean Antarctic Circumpolar Current transport measured in Drake Passage}} {{Mean Antarctic Circumpolar Current transport measured in Drake Passage}}.{\BBCQ}
\newblock
\APACjournalVolNumPages{Geophysical Research Letters}{43}{22}{11--760}.
\newblock
\begin{APACrefDOI} \doi{https://doi.org/10.1002/2016GL070319} \end{APACrefDOI}
\PrintBackRefs{\CurrentBib}

\bibitem [\protect \citeauthoryear {%
Ferrari%
\ \BBA {} Wunsch%
}{%
Ferrari%
\ \BBA {} Wunsch%
}{%
{\protect \APACyear {2009}}%
}]{%
ferrari2009ocean}
\APACinsertmetastar {%
ferrari2009ocean}%
\begin{APACrefauthors}%
Ferrari, R.%
\BCBT {}\ \BBA {} Wunsch, C.%
\end{APACrefauthors}%
\unskip\
\newblock
\APACrefYearMonthDay{2009}{}{}.
\newblock
{\BBOQ}\APACrefatitle {Ocean circulation kinetic energy: Reservoirs, sources, and sinks} {Ocean circulation kinetic energy: Reservoirs, sources, and sinks}.{\BBCQ}
\newblock
\APACjournalVolNumPages{Annual Review of Fluid Mechanics}{41}{}{253--282}.
\newblock
\begin{APACrefDOI} \doi{https://doi.org/10.1146/annurev.fluid.40.111406.102139} \end{APACrefDOI}
\PrintBackRefs{\CurrentBib}

\bibitem [\protect \citeauthoryear {%
Fox-Kemper%
\ \protect \BOthers {.}}{%
Fox-Kemper%
\ \protect \BOthers {.}}{%
{\protect \APACyear {2019}}%
}]{%
fox2019challenges}
\APACinsertmetastar {%
fox2019challenges}%
\begin{APACrefauthors}%
Fox-Kemper, B.%
, Adcroft, A.%
, B{\"o}ning, C\BPBI W.%
, Chassignet, E\BPBI P.%
, Curchitser, E.%
, Danabasoglu, G.%
\BDBL {}others%
\end{APACrefauthors}%
\unskip\
\newblock
\APACrefYearMonthDay{2019}{}{}.
\newblock
{\BBOQ}\APACrefatitle {Challenges and prospects in ocean circulation models} {Challenges and prospects in ocean circulation models}.{\BBCQ}
\newblock
\APACjournalVolNumPages{Frontiers in Marine Science}{6}{}{65}.
\newblock
\begin{APACrefDOI} \doi{https://doi.org/10.3389/fmars.2019.00065} \end{APACrefDOI}
\PrintBackRefs{\CurrentBib}

\bibitem [\protect \citeauthoryear {%
Garcia%
, Boyer%
, Baranova%
\BCBL {}\ \protect \BOthers {.}}{%
Garcia%
\ \protect \BOthers {.}}{%
{\protect \APACyear {2019}}%
}]{%
WOA18data}
\APACinsertmetastar {%
WOA18data}%
\begin{APACrefauthors}%
Garcia, H\BPBI E.%
, Boyer, T\BPBI P.%
, Baranova, O\BPBI K.%
\BCBL {}\ \BOthersPeriod {.}\end{APACrefauthors}%
\unskip\
\newblock
\APACrefYearMonthDay{2019}{}{}.
\newblock
\APACrefbtitle {{World Ocean Atlas 2018: Product Documentation [Dataset]}.} {{World Ocean Atlas 2018: Product Documentation [Dataset]}.}
\newblock
\begin{APACrefURL} \url{https://www.ncei.noaa.gov/data/oceans/woa/WOA18/DATA/} \end{APACrefURL}
\newblock
\begin{APACrefDOI} \doi{https://doi.org/10.25923/tzyw-rp36} \end{APACrefDOI}
\PrintBackRefs{\CurrentBib}

\bibitem [\protect \citeauthoryear {%
Griffies%
}{%
Griffies%
}{%
{\protect \APACyear {2018}}%
}]{%
griffies2018fundamentals}
\APACinsertmetastar {%
griffies2018fundamentals}%
\begin{APACrefauthors}%
Griffies, S\BPBI M.%
\end{APACrefauthors}%
\unskip\
\newblock
\APACrefYear{2018}.
\newblock
\APACrefbtitle {Fundamentals of ocean climate models} {Fundamentals of ocean climate models}.
\newblock
\APACaddressPublisher{}{Princeton University press}.
\newblock
\begin{APACrefDOI} \doi{https://doi.org/10.2307/j.ctv301gzg} \end{APACrefDOI}
\PrintBackRefs{\CurrentBib}

\bibitem [\protect \citeauthoryear {%
Griffies%
\ \protect \BOthers {.}}{%
Griffies%
\ \protect \BOthers {.}}{%
{\protect \APACyear {2024}}%
}]{%
griffies2024cm4xpart1}
\APACinsertmetastar {%
griffies2024cm4xpart1}%
\begin{APACrefauthors}%
Griffies, S\BPBI M.%
, Adcroft, A.%
, Beadling, R\BPBI L.%
, Bushuk, M.%
, Chang, C\BHBI Y.%
, Drake, H\BPBI F.%
\BDBL {}others%
\end{APACrefauthors}%
\unskip\
\newblock
\APACrefYearMonthDay{2024}{}{}.
\newblock
{\BBOQ}\APACrefatitle {{The GFDL-CM4X climate model hierarchy, Part I: model description and thermal properties}} {{The GFDL-CM4X climate model hierarchy, Part I: model description and thermal properties}}.{\BBCQ}
\newblock
\APACjournalVolNumPages{Authorea Preprints}{}{}{}.
\newblock
\begin{APACrefDOI} \doi{https://doi.org/10.22541/essoar.173282145.53065190/v1} \end{APACrefDOI}
\PrintBackRefs{\CurrentBib}

\bibitem [\protect \citeauthoryear {%
Griffies%
\ \protect \BOthers {.}}{%
Griffies%
\ \protect \BOthers {.}}{%
{\protect \APACyear {2015}}%
}]{%
griffies2015impacts}
\APACinsertmetastar {%
griffies2015impacts}%
\begin{APACrefauthors}%
Griffies, S\BPBI M.%
, Winton, M.%
, Anderson, W\BPBI G.%
, Benson, R.%
, Delworth, T\BPBI L.%
, Dufour, C\BPBI O.%
\BDBL {}others%
\end{APACrefauthors}%
\unskip\
\newblock
\APACrefYearMonthDay{2015}{}{}.
\newblock
{\BBOQ}\APACrefatitle {Impacts on ocean heat from transient mesoscale eddies in a hierarchy of climate models} {Impacts on ocean heat from transient mesoscale eddies in a hierarchy of climate models}.{\BBCQ}
\newblock
\APACjournalVolNumPages{Journal of Climate}{28}{3}{952--977}.
\newblock
\begin{APACrefDOI} \doi{https://doi.org/10.1175/JCLI-D-14-00353.1} \end{APACrefDOI}
\PrintBackRefs{\CurrentBib}

\bibitem [\protect \citeauthoryear {%
Grooms%
, Agarwal%
, Marques%
, Pegion%
\BCBL {}\ \BBA {} Yassin%
}{%
Grooms%
\ \protect \BOthers {.}}{%
{\protect \APACyear {2024}}%
}]{%
grooms2024stochastic}
\APACinsertmetastar {%
grooms2024stochastic}%
\begin{APACrefauthors}%
Grooms, I.%
, Agarwal, N.%
, Marques, G\BPBI M.%
, Pegion, P.%
\BCBL {}\ \BBA {} Yassin, H.%
\end{APACrefauthors}%
\unskip\
\newblock
\APACrefYearMonthDay{2024}{}{}.
\newblock
{\BBOQ}\APACrefatitle {{The Stochastic GM+ E closure: A framework for coupling stochastic backscatter with the Gent and McWilliams parameterization}} {{The Stochastic GM+ E closure: A framework for coupling stochastic backscatter with the Gent and McWilliams parameterization}}.{\BBCQ}
\newblock
\APACjournalVolNumPages{Authorea Preprints}{}{}{}.
\newblock
\begin{APACrefDOI} \doi{https://doi.org/10.22541/essoar.172118408.85625257/v1} \end{APACrefDOI}
\PrintBackRefs{\CurrentBib}

\bibitem [\protect \citeauthoryear {%
Grooms%
\ \protect \BOthers {.}}{%
Grooms%
\ \protect \BOthers {.}}{%
{\protect \APACyear {2021}}%
}]{%
grooms2021diffusion}
\APACinsertmetastar {%
grooms2021diffusion}%
\begin{APACrefauthors}%
Grooms, I.%
, Loose, N.%
, Abernathey, R.%
, Steinberg, J.%
, Bachman, S\BPBI D.%
, Marques, G\BPBI M.%
\BDBL {}Yankovsky, E.%
\end{APACrefauthors}%
\unskip\
\newblock
\APACrefYearMonthDay{2021}{}{}.
\newblock
{\BBOQ}\APACrefatitle {Diffusion-based smoothers for spatial filtering of gridded geophysical data} {Diffusion-based smoothers for spatial filtering of gridded geophysical data}.{\BBCQ}
\newblock
\APACjournalVolNumPages{Journal of Advances in Modeling Earth Systems}{13}{9}{e2021MS002552}.
\newblock
\begin{APACrefDOI} \doi{https://doi.org/10.1029/2021MS002552} \end{APACrefDOI}
\PrintBackRefs{\CurrentBib}

\bibitem [\protect \citeauthoryear {%
Guan%
, Subel%
, Chattopadhyay%
\BCBL {}\ \BBA {} Hassanzadeh%
}{%
Guan%
\ \protect \BOthers {.}}{%
{\protect \APACyear {2022}}%
}]{%
guan2022learning}
\APACinsertmetastar {%
guan2022learning}%
\begin{APACrefauthors}%
Guan, Y.%
, Subel, A.%
, Chattopadhyay, A.%
\BCBL {}\ \BBA {} Hassanzadeh, P.%
\end{APACrefauthors}%
\unskip\
\newblock
\APACrefYearMonthDay{2022}{}{}.
\newblock
{\BBOQ}\APACrefatitle {{Learning physics-constrained subgrid-scale closures in the small-data regime for stable and accurate LES}} {{Learning physics-constrained subgrid-scale closures in the small-data regime for stable and accurate LES}}.{\BBCQ}
\newblock
\APACjournalVolNumPages{Physica D: Nonlinear Phenomena}{}{}{133568}.
\newblock
\begin{APACrefDOI} \doi{https://doi.org/10.1016/j.physd.2022.133568} \end{APACrefDOI}
\PrintBackRefs{\CurrentBib}

\bibitem [\protect \citeauthoryear {%
Guillaumin%
\ \BBA {} Zanna%
}{%
Guillaumin%
\ \BBA {} Zanna%
}{%
{\protect \APACyear {2021}}%
}]{%
guillaumin2021stochastic}
\APACinsertmetastar {%
guillaumin2021stochastic}%
\begin{APACrefauthors}%
Guillaumin, A\BPBI P.%
\BCBT {}\ \BBA {} Zanna, L.%
\end{APACrefauthors}%
\unskip\
\newblock
\APACrefYearMonthDay{2021}{}{}.
\newblock
{\BBOQ}\APACrefatitle {Stochastic-deep learning parameterization of ocean momentum forcing} {Stochastic-deep learning parameterization of ocean momentum forcing}.{\BBCQ}
\newblock
\APACjournalVolNumPages{Journal of Advances in Modeling Earth Systems}{13}{9}{e2021MS002534}.
\newblock
\begin{APACrefDOI} \doi{https://doi.org/10.1029/2021MS002534} \end{APACrefDOI}
\PrintBackRefs{\CurrentBib}

\bibitem [\protect \citeauthoryear {%
Gultekin%
\ \protect \BOthers {.}}{%
Gultekin%
\ \protect \BOthers {.}}{%
{\protect \APACyear {2024}}%
}]{%
gultekin2024analysis}
\APACinsertmetastar {%
gultekin2024analysis}%
\begin{APACrefauthors}%
Gultekin, C.%
, Subel, A.%
, Zhang, C.%
, Leibovich, M.%
, Perezhogin, P.%
, Adcroft, A.%
\BDBL {}Zanna, L.%
\end{APACrefauthors}%
\unskip\
\newblock
\APACrefYearMonthDay{2024}{}{}.
\newblock
{\BBOQ}\APACrefatitle {An Analysis of Deep Learning Parameterizations for Ocean Subgrid Eddy Forcing} {An analysis of deep learning parameterizations for ocean subgrid eddy forcing}.{\BBCQ}
\newblock
\APACjournalVolNumPages{arXiv preprint arXiv:2411.06604}{}{}{}.
\newblock
\begin{APACrefDOI} \doi{https://doi.org/10.48550/arXiv.2411.06604} \end{APACrefDOI}
\PrintBackRefs{\CurrentBib}

\bibitem [\protect \citeauthoryear {%
Hastie%
, Tibshirani%
, Friedman%
\BCBL {}\ \BBA {} Friedman%
}{%
Hastie%
\ \protect \BOthers {.}}{%
{\protect \APACyear {2009}}%
}]{%
hastie2009elements}
\APACinsertmetastar {%
hastie2009elements}%
\begin{APACrefauthors}%
Hastie, T.%
, Tibshirani, R.%
, Friedman, J\BPBI H.%
\BCBL {}\ \BBA {} Friedman, J\BPBI H.%
\end{APACrefauthors}%
\unskip\
\newblock
\APACrefYear{2009}.
\newblock
\APACrefbtitle {The elements of statistical learning: data mining, inference, and prediction} {The elements of statistical learning: data mining, inference, and prediction}\ (\BVOL~2).
\newblock
\APACaddressPublisher{}{Springer}.
\PrintBackRefs{\CurrentBib}

\bibitem [\protect \citeauthoryear {%
Hewitt%
\ \protect \BOthers {.}}{%
Hewitt%
\ \protect \BOthers {.}}{%
{\protect \APACyear {2020}}%
}]{%
hewitt2020resolving}
\APACinsertmetastar {%
hewitt2020resolving}%
\begin{APACrefauthors}%
Hewitt, H\BPBI T.%
, Roberts, M.%
, Mathiot, P.%
, Biastoch, A.%
, Blockley, E.%
, Chassignet, E\BPBI P.%
\BDBL {}others%
\end{APACrefauthors}%
\unskip\
\newblock
\APACrefYearMonthDay{2020}{}{}.
\newblock
{\BBOQ}\APACrefatitle {Resolving and parameterising the ocean mesoscale in earth system models} {Resolving and parameterising the ocean mesoscale in earth system models}.{\BBCQ}
\newblock
\APACjournalVolNumPages{Current Climate Change Reports}{6}{4}{137--152}.
\newblock
\begin{APACrefDOI} \doi{https://doi.org/10.1007/s40641-020-00164-w} \end{APACrefDOI}
\PrintBackRefs{\CurrentBib}

\bibitem [\protect \citeauthoryear {%
Jansen%
, Adcroft%
, Khani%
\BCBL {}\ \BBA {} Kong%
}{%
Jansen%
\ \protect \BOthers {.}}{%
{\protect \APACyear {2019}}%
}]{%
jansen2019toward}
\APACinsertmetastar {%
jansen2019toward}%
\begin{APACrefauthors}%
Jansen, M\BPBI F.%
, Adcroft, A.%
, Khani, S.%
\BCBL {}\ \BBA {} Kong, H.%
\end{APACrefauthors}%
\unskip\
\newblock
\APACrefYearMonthDay{2019}{}{}.
\newblock
{\BBOQ}\APACrefatitle {Toward an energetically consistent, resolution aware parameterization of ocean mesoscale eddies} {Toward an energetically consistent, resolution aware parameterization of ocean mesoscale eddies}.{\BBCQ}
\newblock
\APACjournalVolNumPages{Journal of Advances in Modeling Earth Systems}{11}{8}{2844--2860}.
\newblock
\begin{APACrefDOI} \doi{https://doi.org/10.1029/2019MS001750} \end{APACrefDOI}
\PrintBackRefs{\CurrentBib}

\bibitem [\protect \citeauthoryear {%
Jansen%
\ \BBA {} Held%
}{%
Jansen%
\ \BBA {} Held%
}{%
{\protect \APACyear {2014}}%
}]{%
jansen2014parameterizing}
\APACinsertmetastar {%
jansen2014parameterizing}%
\begin{APACrefauthors}%
Jansen, M\BPBI F.%
\BCBT {}\ \BBA {} Held, I\BPBI M.%
\end{APACrefauthors}%
\unskip\
\newblock
\APACrefYearMonthDay{2014}{}{}.
\newblock
{\BBOQ}\APACrefatitle {Parameterizing subgrid-scale eddy effects using energetically consistent backscatter} {Parameterizing subgrid-scale eddy effects using energetically consistent backscatter}.{\BBCQ}
\newblock
\APACjournalVolNumPages{Ocean Modelling}{80}{}{36--48}.
\newblock
\begin{APACrefDOI} \doi{https://doi.org/10.1016/j.ocemod.2014.06.002} \end{APACrefDOI}
\PrintBackRefs{\CurrentBib}

\bibitem [\protect \citeauthoryear {%
Jansen%
, Held%
, Adcroft%
\BCBL {}\ \BBA {} Hallberg%
}{%
Jansen%
\ \protect \BOthers {.}}{%
{\protect \APACyear {2015}}%
}]{%
jansen2015energy}
\APACinsertmetastar {%
jansen2015energy}%
\begin{APACrefauthors}%
Jansen, M\BPBI F.%
, Held, I\BPBI M.%
, Adcroft, A.%
\BCBL {}\ \BBA {} Hallberg, R.%
\end{APACrefauthors}%
\unskip\
\newblock
\APACrefYearMonthDay{2015}{}{}.
\newblock
{\BBOQ}\APACrefatitle {Energy budget-based backscatter in an eddy permitting primitive equation model} {Energy budget-based backscatter in an eddy permitting primitive equation model}.{\BBCQ}
\newblock
\APACjournalVolNumPages{Ocean Modelling}{94}{}{15--26}.
\newblock
\begin{APACrefDOI} \doi{https://doi.org/10.1016/j.ocemod.2015.07.015} \end{APACrefDOI}
\PrintBackRefs{\CurrentBib}

\bibitem [\protect \citeauthoryear {%
Juricke%
, Danilov%
, Kutsenko%
\BCBL {}\ \BBA {} Oliver%
}{%
Juricke%
\ \protect \BOthers {.}}{%
{\protect \APACyear {2019}}%
}]{%
juricke2019ocean}
\APACinsertmetastar {%
juricke2019ocean}%
\begin{APACrefauthors}%
Juricke, S.%
, Danilov, S.%
, Kutsenko, A.%
\BCBL {}\ \BBA {} Oliver, M.%
\end{APACrefauthors}%
\unskip\
\newblock
\APACrefYearMonthDay{2019}{}{}.
\newblock
{\BBOQ}\APACrefatitle {Ocean kinetic energy backscatter parametrizations on unstructured grids: Impact on mesoscale turbulence in a channel} {Ocean kinetic energy backscatter parametrizations on unstructured grids: Impact on mesoscale turbulence in a channel}.{\BBCQ}
\newblock
\APACjournalVolNumPages{Ocean Modelling}{138}{}{51--67}.
\newblock
\begin{APACrefDOI} \doi{https://doi.org/10.1016/j.ocemod.2019.03.009} \end{APACrefDOI}
\PrintBackRefs{\CurrentBib}

\bibitem [\protect \citeauthoryear {%
Kang%
, Jeon%
\BCBL {}\ \BBA {} You%
}{%
Kang%
\ \protect \BOthers {.}}{%
{\protect \APACyear {2023}}%
}]{%
kang2023neural}
\APACinsertmetastar {%
kang2023neural}%
\begin{APACrefauthors}%
Kang, M.%
, Jeon, Y.%
\BCBL {}\ \BBA {} You, D.%
\end{APACrefauthors}%
\unskip\
\newblock
\APACrefYearMonthDay{2023}{}{}.
\newblock
{\BBOQ}\APACrefatitle {Neural-network-based mixed subgrid-scale model for turbulent flow} {Neural-network-based mixed subgrid-scale model for turbulent flow}.{\BBCQ}
\newblock
\APACjournalVolNumPages{Journal of Fluid Mechanics}{962}{}{A38}.
\newblock
\begin{APACrefDOI} \doi{https://doi.org/10.1017/jfm.2023.260} \end{APACrefDOI}
\PrintBackRefs{\CurrentBib}

\bibitem [\protect \citeauthoryear {%
Kashinath%
\ \protect \BOthers {.}}{%
Kashinath%
\ \protect \BOthers {.}}{%
{\protect \APACyear {2021}}%
}]{%
kashinath2021physics}
\APACinsertmetastar {%
kashinath2021physics}%
\begin{APACrefauthors}%
Kashinath, K.%
, Mustafa, M.%
, Albert, A.%
, Wu, J.%
, Jiang, C.%
, Esmaeilzadeh, S.%
\BDBL {}others%
\end{APACrefauthors}%
\unskip\
\newblock
\APACrefYearMonthDay{2021}{}{}.
\newblock
{\BBOQ}\APACrefatitle {Physics-informed machine learning: case studies for weather and climate modelling} {Physics-informed machine learning: case studies for weather and climate modelling}.{\BBCQ}
\newblock
\APACjournalVolNumPages{Philosophical Transactions of the Royal Society A}{379}{2194}{20200093}.
\newblock
\begin{APACrefDOI} \doi{https://doi.org/10.1098/rsta.2020.0093} \end{APACrefDOI}
\PrintBackRefs{\CurrentBib}

\bibitem [\protect \citeauthoryear {%
Large%
\ \BBA {} Yeager%
}{%
Large%
\ \BBA {} Yeager%
}{%
{\protect \APACyear {2009}}%
}]{%
large2009global}
\APACinsertmetastar {%
large2009global}%
\begin{APACrefauthors}%
Large, W.%
\BCBT {}\ \BBA {} Yeager, S.%
\end{APACrefauthors}%
\unskip\
\newblock
\APACrefYearMonthDay{2009}{}{}.
\newblock
{\BBOQ}\APACrefatitle {The global climatology of an interannually varying air--sea flux data set} {The global climatology of an interannually varying air--sea flux data set}.{\BBCQ}
\newblock
\APACjournalVolNumPages{Climate dynamics}{33}{}{341--364}.
\newblock
\begin{APACrefDOI} \doi{https://doi.org/10.1007/s00382-008-0441-3} \end{APACrefDOI}
\PrintBackRefs{\CurrentBib}

\bibitem [\protect \citeauthoryear {%
Leith%
}{%
Leith%
}{%
{\protect \APACyear {1996}}%
}]{%
leith1996stochastic}
\APACinsertmetastar {%
leith1996stochastic}%
\begin{APACrefauthors}%
Leith, C.%
\end{APACrefauthors}%
\unskip\
\newblock
\APACrefYearMonthDay{1996}{}{}.
\newblock
{\BBOQ}\APACrefatitle {Stochastic models of chaotic systems} {Stochastic models of chaotic systems}.{\BBCQ}
\newblock
\APACjournalVolNumPages{Physica D: Nonlinear Phenomena}{98}{2-4}{481--491}.
\newblock
\begin{APACrefDOI} \doi{https://doi.org/10.1016/0167-2789(96)00107-8} \end{APACrefDOI}
\PrintBackRefs{\CurrentBib}

\bibitem [\protect \citeauthoryear {%
L{\'e}vy%
\ \protect \BOthers {.}}{%
L{\'e}vy%
\ \protect \BOthers {.}}{%
{\protect \APACyear {2010}}%
}]{%
levy2010modifications}
\APACinsertmetastar {%
levy2010modifications}%
\begin{APACrefauthors}%
L{\'e}vy, M.%
, Klein, P.%
, Tr{\'e}guier, A\BHBI M.%
, Iovino, D.%
, Madec, G.%
, Masson, S.%
\BCBL {}\ \BBA {} Takahashi, K.%
\end{APACrefauthors}%
\unskip\
\newblock
\APACrefYearMonthDay{2010}{}{}.
\newblock
{\BBOQ}\APACrefatitle {Modifications of gyre circulation by sub-mesoscale physics} {Modifications of gyre circulation by sub-mesoscale physics}.{\BBCQ}
\newblock
\APACjournalVolNumPages{Ocean Modelling}{34}{1-2}{1--15}.
\newblock
\begin{APACrefDOI} \doi{https://doi.org/10.1016/j.ocemod.2010.04.001} \end{APACrefDOI}
\PrintBackRefs{\CurrentBib}

\bibitem [\protect \citeauthoryear {%
Li%
, Xie%
, Zhang%
, Zhang%
\BCBL {}\ \BBA {} Zhao%
}{%
Li%
\ \protect \BOthers {.}}{%
{\protect \APACyear {2025}}%
}]{%
li2025transformer}
\APACinsertmetastar {%
li2025transformer}%
\begin{APACrefauthors}%
Li, H.%
, Xie, J.%
, Zhang, C.%
, Zhang, Y.%
\BCBL {}\ \BBA {} Zhao, Y.%
\end{APACrefauthors}%
\unskip\
\newblock
\APACrefYearMonthDay{2025}{}{}.
\newblock
{\BBOQ}\APACrefatitle {A transformer-based convolutional method to model inverse cascade in forced two-dimensional turbulence} {A transformer-based convolutional method to model inverse cascade in forced two-dimensional turbulence}.{\BBCQ}
\newblock
\APACjournalVolNumPages{Journal of Computational Physics}{520}{}{113475}.
\newblock
\begin{APACrefDOI} \doi{https://doi.org/10.1016/j.jcp.2024.113475} \end{APACrefDOI}
\PrintBackRefs{\CurrentBib}

\bibitem [\protect \citeauthoryear {%
Ling%
, Kurzawski%
\BCBL {}\ \BBA {} Templeton%
}{%
Ling%
\ \protect \BOthers {.}}{%
{\protect \APACyear {2016}}%
}]{%
ling2016reynolds}
\APACinsertmetastar {%
ling2016reynolds}%
\begin{APACrefauthors}%
Ling, J.%
, Kurzawski, A.%
\BCBL {}\ \BBA {} Templeton, J.%
\end{APACrefauthors}%
\unskip\
\newblock
\APACrefYearMonthDay{2016}{}{}.
\newblock
{\BBOQ}\APACrefatitle {Reynolds averaged turbulence modelling using deep neural networks with embedded invariance} {Reynolds averaged turbulence modelling using deep neural networks with embedded invariance}.{\BBCQ}
\newblock
\APACjournalVolNumPages{Journal of Fluid Mechanics}{807}{}{155--166}.
\newblock
\begin{APACrefDOI} \doi{https://doi.org/10.1017/jfm.2016.615} \end{APACrefDOI}
\PrintBackRefs{\CurrentBib}

\bibitem [\protect \citeauthoryear {%
Locarnini%
\ \protect \BOthers {.}}{%
Locarnini%
\ \protect \BOthers {.}}{%
{\protect \APACyear {2018}}%
}]{%
locarnini2018world}
\APACinsertmetastar {%
locarnini2018world}%
\begin{APACrefauthors}%
Locarnini, M.%
, Mishonov, A.%
, Baranova, O.%
, Boyer, T.%
, Zweng, M.%
, Garcia, H.%
\BDBL {}others%
\end{APACrefauthors}%
\unskip\
\newblock
\APACrefYearMonthDay{2018}{}{}.
\newblock
{\BBOQ}\APACrefatitle {{World ocean atlas 2018, volume 1: Temperature}} {{World ocean atlas 2018, volume 1: Temperature}}.{\BBCQ}
\newblock
\APACjournalVolNumPages{NOAA Atlas NESDIS}{}{}{}.
\PrintBackRefs{\CurrentBib}

\bibitem [\protect \citeauthoryear {%
Loose%
\ \protect \BOthers {.}}{%
Loose%
\ \protect \BOthers {.}}{%
{\protect \APACyear {2022}}%
}]{%
loose2022gcm}
\APACinsertmetastar {%
loose2022gcm}%
\begin{APACrefauthors}%
Loose, N.%
, Abernathey, R.%
, Grooms, I.%
, Busecke, J.%
, Guillaumin, A.%
, Yankovsky, E.%
\BDBL {}others%
\end{APACrefauthors}%
\unskip\
\newblock
\APACrefYearMonthDay{2022}{}{}.
\newblock
{\BBOQ}\APACrefatitle {{GCM-filters: A Python package for diffusion-based spatial filtering of gridded data}} {{GCM-filters: A Python package for diffusion-based spatial filtering of gridded data}}.{\BBCQ}
\newblock
\APACjournalVolNumPages{Journal of Open Source Software}{7}{70}{}.
\PrintBackRefs{\CurrentBib}

\bibitem [\protect \citeauthoryear {%
Loose%
, Bachman%
, Grooms%
\BCBL {}\ \BBA {} Jansen%
}{%
Loose%
, Bachman%
\BCBL {}\ \protect \BOthers {.}}{%
{\protect \APACyear {2023}}%
}]{%
loose2023diagnosing}
\APACinsertmetastar {%
loose2023diagnosing}%
\begin{APACrefauthors}%
Loose, N.%
, Bachman, S.%
, Grooms, I.%
\BCBL {}\ \BBA {} Jansen, M.%
\end{APACrefauthors}%
\unskip\
\newblock
\APACrefYearMonthDay{2023}{}{}.
\newblock
{\BBOQ}\APACrefatitle {Diagnosing scale-dependent energy cycles in a high-resolution isopycnal ocean model} {Diagnosing scale-dependent energy cycles in a high-resolution isopycnal ocean model}.{\BBCQ}
\newblock
\APACjournalVolNumPages{Journal of Physical Oceanography}{53}{1}{157--176}.
\newblock
\begin{APACrefDOI} \doi{https://doi.org/10.1175/JPO-D-22-0083.1} \end{APACrefDOI}
\PrintBackRefs{\CurrentBib}

\bibitem [\protect \citeauthoryear {%
Loose%
, Marques%
\BCBL {}\ \protect \BOthers {.}}{%
Loose%
, Marques%
\BCBL {}\ \protect \BOthers {.}}{%
{\protect \APACyear {2023}}%
}]{%
loose2023comparing}
\APACinsertmetastar {%
loose2023comparing}%
\begin{APACrefauthors}%
Loose, N.%
, Marques, G\BPBI M.%
, Adcroft, A.%
, Bachman, S.%
, Griffies, S\BPBI M.%
, Grooms, I.%
\BDBL {}Jansen, M\BPBI F.%
\end{APACrefauthors}%
\unskip\
\newblock
\APACrefYearMonthDay{2023}{}{}.
\newblock
{\BBOQ}\APACrefatitle {Comparing two parameterizations for the restratification effect of mesoscale eddies in an isopycnal ocean model} {Comparing two parameterizations for the restratification effect of mesoscale eddies in an isopycnal ocean model}.{\BBCQ}
\newblock
\APACjournalVolNumPages{Journal of Advances in Modeling Earth Systems}{15}{12}{e2022MS003518}.
\newblock
\begin{APACrefDOI} \doi{https://doi.org/10.1029/2022MS003518} \end{APACrefDOI}
\PrintBackRefs{\CurrentBib}

\bibitem [\protect \citeauthoryear {%
Lund%
\ \BBA {} Novikov%
}{%
Lund%
\ \BBA {} Novikov%
}{%
{\protect \APACyear {1993}}%
}]{%
lund1993parameterization}
\APACinsertmetastar {%
lund1993parameterization}%
\begin{APACrefauthors}%
Lund, T\BPBI S.%
\BCBT {}\ \BBA {} Novikov, E.%
\end{APACrefauthors}%
\unskip\
\newblock
\APACrefYearMonthDay{1993}{}{}.
\newblock
{\BBOQ}\APACrefatitle {Parameterization of subgrid-scale stress by the velocity gradient tensor} {Parameterization of subgrid-scale stress by the velocity gradient tensor}.{\BBCQ}
\newblock
\APACjournalVolNumPages{Annual Research Briefs, 1992}{}{}{}.
\PrintBackRefs{\CurrentBib}

\bibitem [\protect \citeauthoryear {%
Maddison%
}{%
Maddison%
}{%
{\protect \APACyear {2024}}%
}]{%
maddison2024online}
\APACinsertmetastar {%
maddison2024online}%
\begin{APACrefauthors}%
Maddison, J\BPBI R.%
\end{APACrefauthors}%
\unskip\
\newblock
\APACrefYearMonthDay{2024}{}{}.
\newblock
{\BBOQ}\APACrefatitle {Online learning in idealized ocean gyres} {Online learning in idealized ocean gyres}.{\BBCQ}
\newblock
\APACjournalVolNumPages{arXiv preprint arXiv:2412.06393}{}{}{}.
\newblock
\begin{APACrefDOI} \doi{https://doi.org/10.48550/arXiv.2412.06393} \end{APACrefDOI}
\PrintBackRefs{\CurrentBib}

\bibitem [\protect \citeauthoryear {%
Mana%
\ \BBA {} Zanna%
}{%
Mana%
\ \BBA {} Zanna%
}{%
{\protect \APACyear {2014}}%
}]{%
mana2014toward}
\APACinsertmetastar {%
mana2014toward}%
\begin{APACrefauthors}%
Mana, P\BPBI P.%
\BCBT {}\ \BBA {} Zanna, L.%
\end{APACrefauthors}%
\unskip\
\newblock
\APACrefYearMonthDay{2014}{}{}.
\newblock
{\BBOQ}\APACrefatitle {Toward a stochastic parameterization of ocean mesoscale eddies} {Toward a stochastic parameterization of ocean mesoscale eddies}.{\BBCQ}
\newblock
\APACjournalVolNumPages{Ocean Modelling}{79}{}{1--20}.
\newblock
\begin{APACrefDOI} \doi{https://doi.org/10.1016/j.ocemod.2014.04.002} \end{APACrefDOI}
\PrintBackRefs{\CurrentBib}

\bibitem [\protect \citeauthoryear {%
Marques%
\ \protect \BOthers {.}}{%
Marques%
\ \protect \BOthers {.}}{%
{\protect \APACyear {2022}}%
}]{%
marques2022neverworld2}
\APACinsertmetastar {%
marques2022neverworld2}%
\begin{APACrefauthors}%
Marques, G\BPBI M.%
, Loose, N.%
, Yankovsky, E.%
, Steinberg, J\BPBI M.%
, Chang, C\BHBI Y.%
, Bhamidipati, N.%
\BDBL {}others%
\end{APACrefauthors}%
\unskip\
\newblock
\APACrefYearMonthDay{2022}{}{}.
\newblock
{\BBOQ}\APACrefatitle {{NeverWorld2: An idealized model hierarchy to investigate ocean mesoscale eddies across resolutions}} {{NeverWorld2: An idealized model hierarchy to investigate ocean mesoscale eddies across resolutions}}.{\BBCQ}
\newblock
\APACjournalVolNumPages{Geoscientific Model Development}{15}{17}{6567--6579}.
\newblock
\begin{APACrefDOI} \doi{https://doi.org/10.5194/gmd-15-6567-2022} \end{APACrefDOI}
\PrintBackRefs{\CurrentBib}

\bibitem [\protect \citeauthoryear {%
Marques%
\ \protect \BOthers {.}}{%
Marques%
\ \protect \BOthers {.}}{%
{\protect \APACyear {2022}}%
}]{%
NW2data}
\APACinsertmetastar {%
NW2data}%
\begin{APACrefauthors}%
Marques, G\BPBI M.%
\BCBT {}\ \BOthersPeriod {.}
\end{APACrefauthors}%
\unskip\
\newblock
\APACrefYearMonthDay{2022}{}{}.
\newblock
\APACrefbtitle {{Simulation data in idealized ocean configuration NeverWorld2 [Dataset]}.} {{Simulation data in idealized ocean configuration NeverWorld2 [Dataset]}.}
\newblock
\APACaddressPublisher{}{UCAR/NCAR - CISL - CDP}.
\newblock
\begin{APACrefDOI} \doi{https://doi.org/10.26024/f130-ev71} \end{APACrefDOI}
\PrintBackRefs{\CurrentBib}

\bibitem [\protect \citeauthoryear {%
Maulik%
\ \BBA {} San%
}{%
Maulik%
\ \BBA {} San%
}{%
{\protect \APACyear {2017}}%
}]{%
maulik2017neural}
\APACinsertmetastar {%
maulik2017neural}%
\begin{APACrefauthors}%
Maulik, R.%
\BCBT {}\ \BBA {} San, O.%
\end{APACrefauthors}%
\unskip\
\newblock
\APACrefYearMonthDay{2017}{}{}.
\newblock
{\BBOQ}\APACrefatitle {A neural network approach for the blind deconvolution of turbulent flows} {A neural network approach for the blind deconvolution of turbulent flows}.{\BBCQ}
\newblock
\APACjournalVolNumPages{Journal of Fluid Mechanics}{831}{}{151--181}.
\newblock
\begin{APACrefDOI} \doi{https://doi.org/10.1017/jfm.2017.637} \end{APACrefDOI}
\PrintBackRefs{\CurrentBib}

\bibitem [\protect \citeauthoryear {%
Maulik%
, San%
, Rasheed%
\BCBL {}\ \BBA {} Vedula%
}{%
Maulik%
\ \protect \BOthers {.}}{%
{\protect \APACyear {2019}}%
}]{%
maulik2019subgrid}
\APACinsertmetastar {%
maulik2019subgrid}%
\begin{APACrefauthors}%
Maulik, R.%
, San, O.%
, Rasheed, A.%
\BCBL {}\ \BBA {} Vedula, P.%
\end{APACrefauthors}%
\unskip\
\newblock
\APACrefYearMonthDay{2019}{}{}.
\newblock
{\BBOQ}\APACrefatitle {Subgrid modelling for two-dimensional turbulence using neural networks} {Subgrid modelling for two-dimensional turbulence using neural networks}.{\BBCQ}
\newblock
\APACjournalVolNumPages{Journal of Fluid Mechanics}{858}{}{122--144}.
\newblock
\begin{APACrefDOI} \doi{https://doi.org/10.1017/jfm.2018.770} \end{APACrefDOI}
\PrintBackRefs{\CurrentBib}

\bibitem [\protect \citeauthoryear {%
Meneveau%
\ \BBA {} Katz%
}{%
Meneveau%
\ \BBA {} Katz%
}{%
{\protect \APACyear {2000}}%
}]{%
meneveau2000scale}
\APACinsertmetastar {%
meneveau2000scale}%
\begin{APACrefauthors}%
Meneveau, C.%
\BCBT {}\ \BBA {} Katz, J.%
\end{APACrefauthors}%
\unskip\
\newblock
\APACrefYearMonthDay{2000}{}{}.
\newblock
{\BBOQ}\APACrefatitle {Scale-invariance and turbulence models for large-eddy simulation} {Scale-invariance and turbulence models for large-eddy simulation}.{\BBCQ}
\newblock
\APACjournalVolNumPages{Annual Review of Fluid Mechanics}{32}{1}{1--32}.
\newblock
\begin{APACrefDOI} \doi{https://doi.org/10.1146/annurev.fluid.32.1.1} \end{APACrefDOI}
\PrintBackRefs{\CurrentBib}

\bibitem [\protect \citeauthoryear {%
Moat%
, DA%
\BCBL {}\ \protect \BOthers {.}}{%
Moat%
\ \protect \BOthers {.}}{%
{\protect \APACyear {2025}}%
}]{%
RAPIDdata}
\APACinsertmetastar {%
RAPIDdata}%
\begin{APACrefauthors}%
Moat, B.%
, DA, S.%
\BCBL {}\ \BOthersPeriod {.}\end{APACrefauthors}%
\unskip\
\newblock
\APACrefYearMonthDay{2025}{}{}.
\newblock
\APACrefbtitle {{Atlantic meridional overturning circulation observed by the RAPID-MOCHA-WBTS (RAPID-Meridional Overturning Circulation and Heatflux Array-Western Boundary Time Series) array at 26N from 2004 to 2023 (v2023.1a) [Dataset]}.} {{Atlantic meridional overturning circulation observed by the RAPID-MOCHA-WBTS (RAPID-Meridional Overturning Circulation and Heatflux Array-Western Boundary Time Series) array at 26N from 2004 to 2023 (v2023.1a) [Dataset]}.}
\newblock
\APACaddressPublisher{}{British Oceanographic Data Centre - Natural Environment Research Council, UK}.
\newblock
\begin{APACrefDOI} \doi{https://doi.org/10.5285/33826d6e-801c-b0a7-e063-7086abc0b9db} \end{APACrefDOI}
\PrintBackRefs{\CurrentBib}

\bibitem [\protect \citeauthoryear {%
Pawar%
, San%
, Rasheed%
\BCBL {}\ \BBA {} Vedula%
}{%
Pawar%
\ \protect \BOthers {.}}{%
{\protect \APACyear {2020}}%
}]{%
pawar2020priori}
\APACinsertmetastar {%
pawar2020priori}%
\begin{APACrefauthors}%
Pawar, S.%
, San, O.%
, Rasheed, A.%
\BCBL {}\ \BBA {} Vedula, P.%
\end{APACrefauthors}%
\unskip\
\newblock
\APACrefYearMonthDay{2020}{}{}.
\newblock
{\BBOQ}\APACrefatitle {A priori analysis on deep learning of subgrid-scale parameterizations for Kraichnan turbulence} {A priori analysis on deep learning of subgrid-scale parameterizations for kraichnan turbulence}.{\BBCQ}
\newblock
\APACjournalVolNumPages{Theoretical and Computational Fluid Dynamics}{34}{4}{429--455}.
\newblock
\begin{APACrefDOI} \doi{https://doi.org/10.1007/s00162-019-00512-z} \end{APACrefDOI}
\PrintBackRefs{\CurrentBib}

\bibitem [\protect \citeauthoryear {%
Perezhogin%
}{%
Perezhogin%
}{%
{\protect \APACyear {2025}}%
}]{%
perezhogin_2025_software}
\APACinsertmetastar {%
perezhogin_2025_software}%
\begin{APACrefauthors}%
Perezhogin, P.%
\end{APACrefauthors}%
\unskip\
\newblock
\APACrefYearMonthDay{2025}{}{}.
\newblock
\APACrefbtitle {{Generalizable neural-network parameterization of mesoscale eddies in idealized and global ocean models [Software]}.} {{Generalizable neural-network parameterization of mesoscale eddies in idealized and global ocean models [Software]}.}
\newblock
\APACaddressPublisher{}{Zenodo}.
\newblock
\begin{APACrefDOI} \doi{https://doi.org/10.5281/zenodo.16056926} \end{APACrefDOI}
\PrintBackRefs{\CurrentBib}

\bibitem [\protect \citeauthoryear {%
Perezhogin%
, Adcroft%
\BCBL {}\ \BBA {} Zanna%
}{%
Perezhogin%
\ \protect \BOthers {.}}{%
{\protect \APACyear {2025}}%
}]{%
perezhogin_2025_dataset}
\APACinsertmetastar {%
perezhogin_2025_dataset}%
\begin{APACrefauthors}%
Perezhogin, P.%
, Adcroft, A.%
\BCBL {}\ \BBA {} Zanna, L.%
\end{APACrefauthors}%
\unskip\
\newblock
\APACrefYearMonthDay{2025}{}{}.
\newblock
\APACrefbtitle {{Generalizable neural-network parameterization of mesoscale eddies in idealized and global ocean models [Dataset] }.} {{Generalizable neural-network parameterization of mesoscale eddies in idealized and global ocean models [Dataset] }.}
\newblock
\APACaddressPublisher{}{Zenodo}.
\newblock
\begin{APACrefDOI} \doi{https://doi.org/10.5281/zenodo.16058005} \end{APACrefDOI}
\PrintBackRefs{\CurrentBib}

\bibitem [\protect \citeauthoryear {%
Perezhogin%
, Balakrishna%
\BCBL {}\ \BBA {} Agrawal%
}{%
Perezhogin%
, Balakrishna%
\BCBL {}\ \BBA {} Agrawal%
}{%
{\protect \APACyear {2024}}%
}]{%
perezhogin2025large}
\APACinsertmetastar {%
perezhogin2025large}%
\begin{APACrefauthors}%
Perezhogin, P.%
, Balakrishna, A.%
\BCBL {}\ \BBA {} Agrawal, R.%
\end{APACrefauthors}%
\unskip\
\newblock
\APACrefYearMonthDay{2024}{}{}.
\newblock
{\BBOQ}\APACrefatitle {Large eddy simulation of ocean mesoscale eddies} {Large eddy simulation of ocean mesoscale eddies}.{\BBCQ}
\newblock
\BIn{} \APACrefbtitle {{Proceedings of the Summer Program 2024, Center for Turbulence Research, Stanford University}} {{Proceedings of the Summer Program 2024, Center for Turbulence Research, Stanford University}}\ (\BPG~507-516).
\newblock
\begin{APACrefURL} \url{https://arxiv.org/abs/2501.05357} \end{APACrefURL}
\PrintBackRefs{\CurrentBib}

\bibitem [\protect \citeauthoryear {%
Perezhogin%
, Zhang%
, Adcroft%
, Fernandez-Granda%
\BCBL {}\ \BBA {} Zanna%
}{%
Perezhogin%
, Zhang%
\BCBL {}\ \protect \BOthers {.}}{%
{\protect \APACyear {2024}}%
}]{%
perezhogin2024stable}
\APACinsertmetastar {%
perezhogin2024stable}%
\begin{APACrefauthors}%
Perezhogin, P.%
, Zhang, C.%
, Adcroft, A.%
, Fernandez-Granda, C.%
\BCBL {}\ \BBA {} Zanna, L.%
\end{APACrefauthors}%
\unskip\
\newblock
\APACrefYearMonthDay{2024}{}{}.
\newblock
{\BBOQ}\APACrefatitle {A stable implementation of a data-driven scale-aware mesoscale parameterization} {A stable implementation of a data-driven scale-aware mesoscale parameterization}.{\BBCQ}
\newblock
\APACjournalVolNumPages{Journal of Advances in Modeling Earth Systems}{16}{10}{e2023MS004104}.
\newblock
\begin{APACrefDOI} \doi{https://doi.org/10.1029/2023MS004104} \end{APACrefDOI}
\PrintBackRefs{\CurrentBib}

\bibitem [\protect \citeauthoryear {%
Pope%
}{%
Pope%
}{%
{\protect \APACyear {1975}}%
}]{%
pope1975more}
\APACinsertmetastar {%
pope1975more}%
\begin{APACrefauthors}%
Pope, S\BPBI B.%
\end{APACrefauthors}%
\unskip\
\newblock
\APACrefYearMonthDay{1975}{}{}.
\newblock
{\BBOQ}\APACrefatitle {A more general effective-viscosity hypothesis} {A more general effective-viscosity hypothesis}.{\BBCQ}
\newblock
\APACjournalVolNumPages{Journal of Fluid Mechanics}{72}{2}{331--340}.
\newblock
\begin{APACrefDOI} \doi{https://doi.org/10.1017/S0022112075003382} \end{APACrefDOI}
\PrintBackRefs{\CurrentBib}

\bibitem [\protect \citeauthoryear {%
Prakash%
, Jansen%
\BCBL {}\ \BBA {} Evans%
}{%
Prakash%
\ \protect \BOthers {.}}{%
{\protect \APACyear {2022}}%
}]{%
prakash2022invariant}
\APACinsertmetastar {%
prakash2022invariant}%
\begin{APACrefauthors}%
Prakash, A.%
, Jansen, K\BPBI E.%
\BCBL {}\ \BBA {} Evans, J\BPBI A.%
\end{APACrefauthors}%
\unskip\
\newblock
\APACrefYearMonthDay{2022}{}{}.
\newblock
{\BBOQ}\APACrefatitle {Invariant data-driven subgrid stress modeling in the strain-rate eigenframe for large eddy simulation} {Invariant data-driven subgrid stress modeling in the strain-rate eigenframe for large eddy simulation}.{\BBCQ}
\newblock
\APACjournalVolNumPages{Computer Methods in Applied Mechanics and Engineering}{399}{}{115457}.
\newblock
\begin{APACrefDOI} \doi{https://doi.org/10.1016/j.cma.2022.115457} \end{APACrefDOI}
\PrintBackRefs{\CurrentBib}

\bibitem [\protect \citeauthoryear {%
Prakash%
, Jansen%
\BCBL {}\ \BBA {} Evans%
}{%
Prakash%
\ \protect \BOthers {.}}{%
{\protect \APACyear {2024}}%
}]{%
prakash2024invariant}
\APACinsertmetastar {%
prakash2024invariant}%
\begin{APACrefauthors}%
Prakash, A.%
, Jansen, K\BPBI E.%
\BCBL {}\ \BBA {} Evans, J\BPBI A.%
\end{APACrefauthors}%
\unskip\
\newblock
\APACrefYearMonthDay{2024}{}{}.
\newblock
{\BBOQ}\APACrefatitle {Invariant data-driven subgrid stress modeling on anisotropic grids for large eddy simulation} {Invariant data-driven subgrid stress modeling on anisotropic grids for large eddy simulation}.{\BBCQ}
\newblock
\APACjournalVolNumPages{Computer Methods in Applied Mechanics and Engineering}{422}{}{116807}.
\newblock
\begin{APACrefDOI} \doi{https://doi.org/10.1016/j.cma.2024.116807} \end{APACrefDOI}
\PrintBackRefs{\CurrentBib}

\bibitem [\protect \citeauthoryear {%
Reissmann%
, Hasslberger%
, Sandberg%
\BCBL {}\ \BBA {} Klein%
}{%
Reissmann%
\ \protect \BOthers {.}}{%
{\protect \APACyear {2021}}%
}]{%
reissmann2021application}
\APACinsertmetastar {%
reissmann2021application}%
\begin{APACrefauthors}%
Reissmann, M.%
, Hasslberger, J.%
, Sandberg, R\BPBI D.%
\BCBL {}\ \BBA {} Klein, M.%
\end{APACrefauthors}%
\unskip\
\newblock
\APACrefYearMonthDay{2021}{}{}.
\newblock
{\BBOQ}\APACrefatitle {Application of gene expression programming to a-posteriori LES modeling of a Taylor Green vortex} {Application of gene expression programming to a-posteriori les modeling of a taylor green vortex}.{\BBCQ}
\newblock
\APACjournalVolNumPages{Journal of Computational Physics}{424}{}{109859}.
\newblock
\begin{APACrefDOI} \doi{https://doi.org/10.1016/j.jcp.2020.109859} \end{APACrefDOI}
\PrintBackRefs{\CurrentBib}

\bibitem [\protect \citeauthoryear {%
Ross%
, Li%
, Perezhogin%
, Fernandez-Granda%
\BCBL {}\ \BBA {} Zanna%
}{%
Ross%
\ \protect \BOthers {.}}{%
{\protect \APACyear {2023}}%
}]{%
ross2022benchmarking}
\APACinsertmetastar {%
ross2022benchmarking}%
\begin{APACrefauthors}%
Ross, A.%
, Li, Z.%
, Perezhogin, P.%
, Fernandez-Granda, C.%
\BCBL {}\ \BBA {} Zanna, L.%
\end{APACrefauthors}%
\unskip\
\newblock
\APACrefYearMonthDay{2023}{}{}.
\newblock
{\BBOQ}\APACrefatitle {Benchmarking of machine learning ocean subgrid parameterizations in an idealized model} {Benchmarking of machine learning ocean subgrid parameterizations in an idealized model}.{\BBCQ}
\newblock
\APACjournalVolNumPages{Journal of Advances in Modeling Earth Systems}{15}{1}{e2022MS003258}.
\newblock
\begin{APACrefDOI} \doi{https://doi.org/10.1029/2022MS003258} \end{APACrefDOI}
\PrintBackRefs{\CurrentBib}

\bibitem [\protect \citeauthoryear {%
Salmon%
}{%
Salmon%
}{%
{\protect \APACyear {1980}}%
}]{%
salmon1980baroclinic}
\APACinsertmetastar {%
salmon1980baroclinic}%
\begin{APACrefauthors}%
Salmon, R.%
\end{APACrefauthors}%
\unskip\
\newblock
\APACrefYearMonthDay{1980}{}{}.
\newblock
{\BBOQ}\APACrefatitle {Baroclinic instability and geostrophic turbulence} {Baroclinic instability and geostrophic turbulence}.{\BBCQ}
\newblock
\APACjournalVolNumPages{Geophysical \& Astrophysical Fluid Dynamics}{15}{1}{167--211}.
\newblock
\begin{APACrefDOI} \doi{https://doi.org/10.1080/03091928008241178} \end{APACrefDOI}
\PrintBackRefs{\CurrentBib}

\bibitem [\protect \citeauthoryear {%
Sane%
, Reichl%
, Adcroft%
\BCBL {}\ \BBA {} Zanna%
}{%
Sane%
\ \protect \BOthers {.}}{%
{\protect \APACyear {2023}}%
}]{%
Sane2023}
\APACinsertmetastar {%
Sane2023}%
\begin{APACrefauthors}%
Sane, A.%
, Reichl, B\BPBI G.%
, Adcroft, A.%
\BCBL {}\ \BBA {} Zanna, L.%
\end{APACrefauthors}%
\unskip\
\newblock
\APACrefYearMonthDay{2023}{}{}.
\newblock
{\BBOQ}\APACrefatitle {Parameterizing Vertical Mixing Coefficients in the Ocean Surface Boundary Layer Using Neural Networks} {Parameterizing vertical mixing coefficients in the ocean surface boundary layer using neural networks}.{\BBCQ}
\newblock
\APACjournalVolNumPages{Journal of Advances in Modeling Earth Systems}{15}{10}{e2023MS003890}.
\newblock
\begin{APACrefDOI} \doi{https://doi.org/10.1029/2023MS003890} \end{APACrefDOI}
\PrintBackRefs{\CurrentBib}

\bibitem [\protect \citeauthoryear {%
Schneider%
, Leung%
\BCBL {}\ \BBA {} Wills%
}{%
Schneider%
\ \protect \BOthers {.}}{%
{\protect \APACyear {2024}}%
}]{%
schneider2024opinion}
\APACinsertmetastar {%
schneider2024opinion}%
\begin{APACrefauthors}%
Schneider, T.%
, Leung, L\BPBI R.%
\BCBL {}\ \BBA {} Wills, R\BPBI C.%
\end{APACrefauthors}%
\unskip\
\newblock
\APACrefYearMonthDay{2024}{}{}.
\newblock
{\BBOQ}\APACrefatitle {Opinion: Optimizing climate models with process knowledge, resolution, and artificial intelligence} {Opinion: Optimizing climate models with process knowledge, resolution, and artificial intelligence}.{\BBCQ}
\newblock
\APACjournalVolNumPages{Atmospheric Chemistry and Physics}{24}{12}{7041--7062}.
\newblock
\begin{APACrefDOI} \doi{https://doi.org/10.5194/acp-24-7041-2024} \end{APACrefDOI}
\PrintBackRefs{\CurrentBib}

\bibitem [\protect \citeauthoryear {%
Shankar%
, Chakraborty%
, Viswanathan%
\BCBL {}\ \BBA {} Maulik%
}{%
Shankar%
\ \protect \BOthers {.}}{%
{\protect \APACyear {2025}}%
}]{%
shankar2025differentiable}
\APACinsertmetastar {%
shankar2025differentiable}%
\begin{APACrefauthors}%
Shankar, V.%
, Chakraborty, D.%
, Viswanathan, V.%
\BCBL {}\ \BBA {} Maulik, R.%
\end{APACrefauthors}%
\unskip\
\newblock
\APACrefYearMonthDay{2025}{}{}.
\newblock
{\BBOQ}\APACrefatitle {Differentiable turbulence: Closure as a partial differential equation constrained optimization} {Differentiable turbulence: Closure as a partial differential equation constrained optimization}.{\BBCQ}
\newblock
\APACjournalVolNumPages{Physical Review Fluids}{10}{2}{024605}.
\PrintBackRefs{\CurrentBib}

\bibitem [\protect \citeauthoryear {%
Smagorinsky%
}{%
Smagorinsky%
}{%
{\protect \APACyear {1963}}%
}]{%
smagorinsky1963general}
\APACinsertmetastar {%
smagorinsky1963general}%
\begin{APACrefauthors}%
Smagorinsky, J.%
\end{APACrefauthors}%
\unskip\
\newblock
\APACrefYearMonthDay{1963}{}{}.
\newblock
{\BBOQ}\APACrefatitle {General circulation experiments with the primitive equations: I. The basic experiment} {General circulation experiments with the primitive equations: I. the basic experiment}.{\BBCQ}
\newblock
\APACjournalVolNumPages{Monthly weather review}{91}{3}{99--164}.
\newblock
\begin{APACrefDOI} \doi{https://doi.org/10.1175/1520-0493(1963)091%3C0099:GCEWTP%3E2.3.CO;2} \end{APACrefDOI}
\PrintBackRefs{\CurrentBib}

\bibitem [\protect \citeauthoryear {%
Srinivasan%
, Chekroun%
\BCBL {}\ \BBA {} McWilliams%
}{%
Srinivasan%
\ \protect \BOthers {.}}{%
{\protect \APACyear {2024}}%
}]{%
srinivasan2024turbulence}
\APACinsertmetastar {%
srinivasan2024turbulence}%
\begin{APACrefauthors}%
Srinivasan, K.%
, Chekroun, M\BPBI D.%
\BCBL {}\ \BBA {} McWilliams, J\BPBI C.%
\end{APACrefauthors}%
\unskip\
\newblock
\APACrefYearMonthDay{2024}{}{}.
\newblock
{\BBOQ}\APACrefatitle {Turbulence closure with small, local neural networks: Forced two-dimensional and $\beta$-plane flows} {Turbulence closure with small, local neural networks: Forced two-dimensional and $\beta$-plane flows}.{\BBCQ}
\newblock
\APACjournalVolNumPages{Journal of Advances in Modeling Earth Systems}{16}{4}{e2023MS003795}.
\newblock
\begin{APACrefDOI} \doi{https://doi.org/10.1029/2023MS003795} \end{APACrefDOI}
\PrintBackRefs{\CurrentBib}

\bibitem [\protect \citeauthoryear {%
Wang%
, Yuan%
, Wang%
\BCBL {}\ \BBA {} Wang%
}{%
Wang%
\ \protect \BOthers {.}}{%
{\protect \APACyear {2022}}%
}]{%
wang2022constant}
\APACinsertmetastar {%
wang2022constant}%
\begin{APACrefauthors}%
Wang, Y.%
, Yuan, Z.%
, Wang, X.%
\BCBL {}\ \BBA {} Wang, J.%
\end{APACrefauthors}%
\unskip\
\newblock
\APACrefYearMonthDay{2022}{}{}.
\newblock
{\BBOQ}\APACrefatitle {Constant-coefficient spatial gradient models for the sub-grid scale closure in large-eddy simulation of turbulence} {Constant-coefficient spatial gradient models for the sub-grid scale closure in large-eddy simulation of turbulence}.{\BBCQ}
\newblock
\APACjournalVolNumPages{Physics of Fluids}{34}{9}{}.
\newblock
\begin{APACrefDOI} \doi{https://doi.org/10.1063/5.0101356} \end{APACrefDOI}
\PrintBackRefs{\CurrentBib}

\bibitem [\protect \citeauthoryear {%
Wang%
, Yuan%
, Xie%
\BCBL {}\ \BBA {} Wang%
}{%
Wang%
\ \protect \BOthers {.}}{%
{\protect \APACyear {2021}}%
}]{%
wang2021artificial}
\APACinsertmetastar {%
wang2021artificial}%
\begin{APACrefauthors}%
Wang, Y.%
, Yuan, Z.%
, Xie, C.%
\BCBL {}\ \BBA {} Wang, J.%
\end{APACrefauthors}%
\unskip\
\newblock
\APACrefYearMonthDay{2021}{}{}.
\newblock
{\BBOQ}\APACrefatitle {Artificial neural network-based spatial gradient models for large-eddy simulation of turbulence} {Artificial neural network-based spatial gradient models for large-eddy simulation of turbulence}.{\BBCQ}
\newblock
\APACjournalVolNumPages{AIP Advances}{11}{5}{}.
\newblock
\begin{APACrefDOI} \doi{https://doi.org/10.1063/5.0053590} \end{APACrefDOI}
\PrintBackRefs{\CurrentBib}

\bibitem [\protect \citeauthoryear {%
Xie%
, Wang%
\BCBL {}\ \BBA {} E%
}{%
Xie%
\ \protect \BOthers {.}}{%
{\protect \APACyear {2020}}%
}]{%
xie2020modeling}
\APACinsertmetastar {%
xie2020modeling}%
\begin{APACrefauthors}%
Xie, C.%
, Wang, J.%
\BCBL {}\ \BBA {} E, W.%
\end{APACrefauthors}%
\unskip\
\newblock
\APACrefYearMonthDay{2020}{}{}.
\newblock
{\BBOQ}\APACrefatitle {Modeling subgrid-scale forces by spatial artificial neural networks in large eddy simulation of turbulence} {Modeling subgrid-scale forces by spatial artificial neural networks in large eddy simulation of turbulence}.{\BBCQ}
\newblock
\APACjournalVolNumPages{Physical Review Fluids}{5}{5}{054606}.
\newblock
\begin{APACrefDOI} \doi{https://doi.org/10.1103/PhysRevFluids.5.054606} \end{APACrefDOI}
\PrintBackRefs{\CurrentBib}

\bibitem [\protect \citeauthoryear {%
Yan%
, Mak%
\BCBL {}\ \BBA {} Wang%
}{%
Yan%
\ \protect \BOthers {.}}{%
{\protect \APACyear {2024}}%
}]{%
yan2024choice}
\APACinsertmetastar {%
yan2024choice}%
\begin{APACrefauthors}%
Yan, F\BPBI E.%
, Mak, J.%
\BCBL {}\ \BBA {} Wang, Y.%
\end{APACrefauthors}%
\unskip\
\newblock
\APACrefYearMonthDay{2024}{}{}.
\newblock
{\BBOQ}\APACrefatitle {On the choice of training data for machine learning of geostrophic mesoscale turbulence} {On the choice of training data for machine learning of geostrophic mesoscale turbulence}.{\BBCQ}
\newblock
\APACjournalVolNumPages{{Journal of Advances in Modeling Earth Systems}}{16}{2}{e2023MS003915}.
\newblock
\begin{APACrefDOI} \doi{https://doi.org/10.1029/2023MS003915} \end{APACrefDOI}
\PrintBackRefs{\CurrentBib}

\bibitem [\protect \citeauthoryear {%
Yankovsky%
, Bachman%
, Smith%
\BCBL {}\ \BBA {} Zanna%
}{%
Yankovsky%
\ \protect \BOthers {.}}{%
{\protect \APACyear {2024}}%
}]{%
yankovsky2024}
\APACinsertmetastar {%
yankovsky2024}%
\begin{APACrefauthors}%
Yankovsky, E.%
, Bachman, S.%
, Smith, K\BPBI S.%
\BCBL {}\ \BBA {} Zanna, L.%
\end{APACrefauthors}%
\unskip\
\newblock
\APACrefYearMonthDay{2024}{}{}.
\newblock
{\BBOQ}\APACrefatitle {Vertical structure and energetic constraints for a backscatter parameterization of ocean mesoscale eddies} {Vertical structure and energetic constraints for a backscatter parameterization of ocean mesoscale eddies}.{\BBCQ}
\newblock
\APACjournalVolNumPages{Journal of Advances in Modeling Earth Systems}{16}{7}{e2023MS004093}.
\newblock
\begin{APACrefDOI} \doi{https://doi.org/10.1029/2023MS004093} \end{APACrefDOI}
\PrintBackRefs{\CurrentBib}

\bibitem [\protect \citeauthoryear {%
Zanna%
\ \BBA {} Bolton%
}{%
Zanna%
\ \BBA {} Bolton%
}{%
{\protect \APACyear {2020}}%
}]{%
zanna2020data}
\APACinsertmetastar {%
zanna2020data}%
\begin{APACrefauthors}%
Zanna, L.%
\BCBT {}\ \BBA {} Bolton, T.%
\end{APACrefauthors}%
\unskip\
\newblock
\APACrefYearMonthDay{2020}{}{}.
\newblock
{\BBOQ}\APACrefatitle {Data-driven equation discovery of ocean mesoscale closures} {Data-driven equation discovery of ocean mesoscale closures}.{\BBCQ}
\newblock
\APACjournalVolNumPages{Geophysical Research Letters}{47}{17}{e2020GL088376}.
\newblock
\begin{APACrefDOI} \doi{https://doi.org/10.1029/2020GL088376} \end{APACrefDOI}
\PrintBackRefs{\CurrentBib}

\bibitem [\protect \citeauthoryear {%
Zhang%
\ \protect \BOthers {.}}{%
Zhang%
\ \protect \BOthers {.}}{%
{\protect \APACyear {2023}}%
}]{%
zhang2023implementation}
\APACinsertmetastar {%
zhang2023implementation}%
\begin{APACrefauthors}%
Zhang, C.%
, Perezhogin, P.%
, Gultekin, C.%
, Adcroft, A.%
, Fernandez-Granda, C.%
\BCBL {}\ \BBA {} Zanna, L.%
\end{APACrefauthors}%
\unskip\
\newblock
\APACrefYearMonthDay{2023}{}{}.
\newblock
{\BBOQ}\APACrefatitle {Implementation and evaluation of a machine learned mesoscale eddy parameterization into a numerical ocean circulation model} {Implementation and evaluation of a machine learned mesoscale eddy parameterization into a numerical ocean circulation model}.{\BBCQ}
\newblock
\APACjournalVolNumPages{Journal of Advances in Modeling Earth Systems}{15}{10}{e2023MS003697}.
\newblock
\begin{APACrefDOI} \doi{https://doi.org/10.1029/2023MS003697} \end{APACrefDOI}
\PrintBackRefs{\CurrentBib}

\end{thebibliography}


%
%
%
%
%

\end{document}


\makeatletter
\def\@makecol{\setbox\@outputbox
     \vbox{\boxmaxdepth \maxdepth
\ifdim\ht\dbltopins<1pt\else\unvbox\dbltopins\fi
     \unvbox\@cclv
\ifdim\ht\dblbotins<1pt\else\unvbox\dblbotins\fi%
\ifvoid\footins\else\vskip\skip\footins\footnoterule\unvbox\footins\fi
\vskip 0pt plus 1fil minus \maxdimen
}%
\global\savedblfigandtabnumber\dblfigandtabnumber
   \xdef\@freelist{\@freelist\@midlist}\gdef\@midlist{}\@combinefloats
   \setbox\@outputbox\vbox to\@colht{\boxmaxdepth\maxdepth
   \@texttop\dimen128=\dp\@outputbox\unvbox\@outputbox
   \vskip-\dimen128\@textbottom}%
   \global\maxdepth\@maxdepth}
\makeatother

\title{Supporting Information for "Generalizable neural-network parameterization of mesoscale eddies in idealized and global ocean models"}

\vspace{0.5 cm}

\authors{Pavel Perezhogin\affil{1}, Alistair Adcroft\affil{3}, Laure Zanna\affil{1,2}}

\affiliation{1}{Courant Institute of Mathematical Sciences, New York University, New York, NY, USA}
\affiliation{2}{Center for Data Science, New York University, New York, NY, USA}

\affiliation{3}{Program in Atmospheric and Oceanic Sciences, Princeton University, Princeton, NJ, USA}

\begin{article}

\noindent\textbf{Contents of this file}
\begin{enumerate}
\item Text S1 to S4
\item Tables S1 to S3
\item Figures S1 to S6
\end{enumerate}



\newpage

\section{*}{Text S1. Known parameterizations as a special case of dimensional scaling}
Here, we show that enforcing the dimensional scaling constraint to the ANN parameterization is not too restrictive and admits multiple known parameterizations as a special case with continuous functional representations (see \citeA{prakash2022invariant} for discussion). We also show that these functional representations do not depend on the normalization factor ($||\mathbf{X}||_2^2$) explicitly. This suggests that the choice of the normalization factor primarily affects the range of inputs to the neural network, but not the function to be learnt. 

We denote the components of the predicted momentum fluxes as follows:
\begin{gather}
    \widehat{\mathbf{T}}(\mathbf{X}, \Delta) = \Delta^2 ||\mathbf{X}||_2^2 \mathrm{ANN}_{\theta} (\mathbf{X} / ||\mathbf{X}||_2) \equiv \\ 
    \Delta^2 ||\mathbf{X}||_2^2 \begin{pmatrix}
        \mathrm{ANN}_{\theta}^{xx} (\mathbf{x}) & \mathrm{ANN}_{\theta}^{xy} (\mathbf{x}) \\
        \mathrm{ANN}_{\theta}^{xy} (\mathbf{x}) & \mathrm{ANN}_{\theta}^{yy} (\mathbf{x})
    \end{pmatrix}, \label{eq:second_ANN_equation}
\end{gather}
where the vector of input features is
\begin{equation}
    \mathbf{X} =
    \begin{pmatrix}
        \left[ \overline{\sigma}_S \right]{\updownarrow \scriptstyle 9}  \\
        \left[ \overline{\sigma}_T \right]{\updownarrow \scriptstyle 9} \\
        \left[ \overline{\omega} \right]{\updownarrow \scriptstyle 9}
    \end{pmatrix} \in \mathbb{R}^{27}
\end{equation}
and $\mathbf{x}=\mathbf{X} / ||\mathbf{X}||_2$.

\subsection{*}{Smagorinsky parameterization}
We first consider a \citeA{smagorinsky1963general} subgrid parameterization:
\begin{equation}
    \widehat{\mathbf{T}} =   C_S \Delta^2 \sqrt{\overline{\sigma}_S^2 + \overline{\sigma}_T^2} \begin{pmatrix}
    \overline{\sigma}_T & \overline{\sigma}_S \\
    \overline{\sigma}_S & -\overline{\sigma}_T
    \end{pmatrix}.
\end{equation}
This subgrid model can be given in the form of Eq. \eqref{eq:second_ANN_equation} if ANN parameterizes the following functions:
\begin{gather}
    \mathrm{ANN}_{\theta}^{xx}(\mathbf{x}) = C_S x_{14} \sqrt{x_5^2 + x_{14}^2} \\
    \mathrm{ANN}_{\theta}^{yy}(\mathbf{x}) = - \mathrm{ANN}_{\theta}^{xx}(\mathbf{x}) \\
    \mathrm{ANN}_{\theta}^{xy}(\mathbf{x}) = C_S x_{5} \sqrt{x_5^2 + x_{14}^2},
\end{gather}
where $x_5$ and $x_{14}$ represent components of the non-dimensional vector $\mathbf{x}$ which are equal to $\overline{\sigma}_S / ||\mathbf{X}||_2$  and $\overline{\sigma}_T / ||\mathbf{X}||_2$ in the center of $3 \times 3$ spatial stencil, respectively. The derived functions are continuous on a bounded domain ($|x_i|\leq 1$), and thus they can be easily learned with the ANN. The functional representations of the parameterizations derived below are continuous as well.

\subsection{*}{Zanna-Bolton 2020 parameterization}
Similarly, we can show that \citeA{zanna2020data} parameterization
\begin{equation}
    \widehat{\mathbf{T}} = - \gamma \Delta^2 
    \begin{pmatrix}
        -\overline{\omega} \, \overline{\sigma}_S & \overline{\omega} \, \overline{\sigma}_T\\
        \overline{\omega} \, \overline{\sigma}_T & \overline{\omega} \, \overline{\sigma}_S
    \end{pmatrix} - \frac{1}{2}\gamma \Delta^2 (
        \overline{\omega}^2 + \overline{\sigma}_T^2 + \overline{\sigma}_S^2)
    \begin{pmatrix}
        1 & 0\\
        0 & 1
    \end{pmatrix}
\end{equation}
can be represented as follows:
\begin{gather}
    \mathrm{ANN}_{\theta}^{xx}(\mathbf{x}) =  \gamma x_5 x_{23} - \frac{1}{2} \gamma (x_5^2 + x_{14}^2 + x_{23}^2), \\
    \mathrm{ANN}_{\theta}^{yy}(\mathbf{x}) =  -\gamma x_5 x_{23} - \frac{1}{2} \gamma (x_5^2 + x_{14}^2 + x_{23}^2), \\
    ANN_{\theta}^{xy}(\mathbf{x}) =  -\gamma x_{14} x_{23}. 
\end{gather}

\subsection{*}{Leith 1996 parameterization} Next, we consider \citeA{leith1996stochastic} parameterization:
\begin{equation}
    \widehat{\mathbf{T}} = C_L \Delta^3 | \nabla \overline{\omega}| \begin{pmatrix}
    \overline{\sigma}_T & \overline{\sigma}_S \\
    \overline{\sigma}_S & -\overline{\sigma}_T
    \end{pmatrix}.
\end{equation}
By approximating the gradient with central differences and assuming an isotropic and uniform grid, we obtain:
\begin{gather}
    \mathrm{ANN}_{\theta}^{xx}(\mathbf{x}) = \frac{1}{2} C_L x_{14} \sqrt{(x_{24}-x_{22})^2 + (x_{26}-x_{20})^2}, \\
    \mathrm{ANN}_{\theta}^{yy}(\mathbf{x}) = -\mathrm{ANN}_{\theta}^{xx}(\mathbf{x}), \\
    \mathrm{ANN}_{\theta}^{xy}(\mathbf{x}) = \frac{1}{2} C_L x_{5} \sqrt{(x_{24}-x_{22})^2 + (x_{26}-x_{20})^2}
\end{gather}

\subsection{*}{Biharmonic Smagorinsky parameterization} The biharmonic Smagorinsky subgrid model has the form:
\begin{equation}
    \widehat{\mathbf{T}} =   -C_S \Delta^4 \sqrt{\overline{\sigma}_S^2 + \overline{\sigma}_T^2} 
    \nabla^2
    \begin{pmatrix}
     \overline{\sigma}_T & \overline{\sigma}_S \\
    \overline{\sigma}_S & -\overline{\sigma}_T
    \end{pmatrix}.
\end{equation}
By approximating the $\nabla^2$ operator on an isotropic and uniform grid, we obtain:
\begin{gather}
    \mathrm{ANN}_{\theta}^{xx}(\mathbf{x}) = - C_S (x_{15} + x_{13} + x_{17} + x_{11}-4x_{14}) \sqrt{x_5^2 + x_{14}^2}, \\
        \mathrm{ANN}_{\theta}^{yy}(\mathbf{x}) = -\mathrm{ANN}_{\theta}^{xx}(\mathbf{x}), \\
      \mathrm{ANN}_{\theta}^{xy}(\mathbf{x}) = - C_S (x_{6} + x_{4} + x_{8} + x_{2}-4x_{5}) \sqrt{x_5^2 + x_{14}^2}.
\end{gather}

\section{*}{Text S2. Robustness of division by small numbers}
The robustness of the  parameterization with the dimensional scaling, 
\begin{gather}
    \widehat{\mathbf{T}}(\mathbf{X}, \Delta) = \Delta^2 ||\mathbf{X}||_2^2 \mathrm{ANN}_{\theta} (\mathbf{X} / ||\mathbf{X}||_2), \label{eq:first_ANN_equation}
\end{gather}
at $\mathbf{X}=\mathbf{0}$ is achieved as follows. We first identify that most known parameterizations, such as \citeA{smagorinsky1963general} and \citeA{leith1996stochastic}, predict zero fluxes when the velocity gradients are zero. We enforce the same property for our parameterization by extending Eq. \eqref{eq:first_ANN_equation} with:
\begin{equation}
    \widehat{\mathbf{T}}(\mathbf{X}=\mathbf{0}, \Delta)=\mathbf{0}.
\end{equation}
Numerically, this property is implemented by adding a very small number ($10^{-30}\mathrm{s}^{-1}$) to the denominator in Eq. \eqref{eq:first_ANN_equation}. 

Additionally, we ensure that Eq. \eqref{eq:first_ANN_equation} is continuous at $\mathbf{X}=\mathbf{0}$, that is, the corresponding limit exists and is equal to the function value (zero):
\begin{equation}
    \lim_{\mathbf{X}\to\mathbf{0}} \widehat{\mathbf{T}}(\mathbf{X}, \Delta) = \widehat{\mathbf{T}}(\mathbf{X}=\mathbf{0}, \Delta) = \mathbf{0}. \label{eq:limit}
\end{equation}
The function $\mathrm{ANN}_{\theta}$ is continuous as a composition of continuous activation functions (ReLU). Furthermore, for any $||\mathbf{X}||_2>0$, the function $\mathrm{ANN}_{\theta}$ is evaluated on a unit sphere, which is a compact set. Therefore, the continuous function $\mathrm{ANN}_{\theta}$ is bounded on the compact set by some constant $A(\theta)$ that depends only on the trainable parameters $\theta$. We verify the limit (Eq. \eqref{eq:limit}) by inequality:
\begin{equation}
    \left|\Delta^2 ||\mathbf{X}||_2^2 \mathrm{ANN}_{\theta} (\mathbf{X} / ||\mathbf{X}||_2)\right| \leq A(\theta) \Delta^2 ||\mathbf{X}||_2^2 \to \mathbf{0} \text{ as } \mathbf{X}\to\mathbf{0}.
\end{equation}

\section{*}{Text S3. Details of the training algorithm} 
The training dataset is created using four coarse-graining factors, selected to be similar to those used in \citeA{gultekin2024analysis}, and 10 depths (extending \citeA{gultekin2024analysis}), see Table \ref{tab:parameters}. 

\subsection{*}{ANN model architecture}
For offline analysis, we use an ANN, also known as a multilayer perceptron (MLP), with two hidden layers, 32 neurons each, in a total of 2051 parameters, both for parameterizations with and without dimensional scaling (see Table \ref{tab:parameters}). For online implementation, the ANN model is chosen to be smaller (see Table \ref{tab:parameters}): it has only a single hidden layer with $20$ neurons, as in \citeA{prakash2022invariant}, with a total of $623$ trainable parameters. We verified that reducing the number of neurons for online implementation does not significantly impact the response in kinetic and potential energy and the time-mean sea surface temperature in short 5-year simulations in the global ocean model OM4 (Figure S5). Thus, we keep the smaller ANN for online implementation to bound its computational cost to within $\approx 10\%$ of the global ocean model runtime.

\subsection{*}{Training algorithm and boundary conditions}
We train the ANN model on data from the full globe, similarly to \citeA{gultekin2024analysis}. The loss function is defined to optimize for the divergence ($\nabla \cdot$ ) of subfilter fluxes (${\mathbf{T}}$) similarly to \citeA{zanna2020data} and \citeA{srinivasan2024turbulence}. The mean squared error (MSE) loss is minimized on every 2D snapshot of subfilter forcing $\mathbf{S}$ and normalized by the corresponding $l_2$-norm of $\mathbf{S}$ \cite{agdestein2025discretize}:
\begin{equation}
\mathcal{L}_{\mathrm{MSE}} = ||(\mathbf{S} - \nabla \cdot \widehat{\mathbf{T}}) \cdot m ||_2^2 ~ / ~||\mathbf{S} \cdot m||_2^2, \label{eq:mse_loss}
\end{equation}
where $m$ is the mask of wet points. The
input features (velocity gradients, $\mathbf{X}$) and predicted subfilter fluxes are set to zero on the land as well: $\widehat{\mathbf{T}}\equiv m \cdot \widehat{\mathbf{T}}(m \cdot \mathbf{X})$. That is, we impose zero Neumann boundary condition \cite{zhang2024addressing} and free-slip boundary condition. We found that including the grid points adjacent to the land to the loss function is essential for ensuring the numerical stability of online runs. Another important design choice for online numerical stability is performing the ANN inference on the collocated, rather than on the staggered grid, similarly to \citeA{guillaumin2021stochastic} and  \citeA{agdestein2025discretize}. 

The loss function (Eq. \eqref{eq:mse_loss}) is evaluated and minimized for a total of $16000$ two-dimensional snapshots during training, see Table \ref{tab:parameters}. We do not use any regularizations, such as weight decay, during the training of ANNs because the size of the dataset is much bigger relative to the number of trainable parameters. We verified that the offline skill on training and testing data is very similar, suggesting that there is no overfitting, and there is no need for regularization.

\subsection{*}{Sensitivity to the random seed}
The R-squared of the offline predictions of the ANN is almost insensitive to the random seed used to initialize the training algorithm. In addition, the prediction errors $\mathbf{S} - \nabla \cdot \widehat{\mathbf{T}}$ are highly correlated between different seeds (as in \citeA{srinivasan2024turbulence}). We also confirmed that the kinetic energy is nearly unchanged in online two-layer Double Gyre experiments, using ANNs generated from different initializations of the training algorithm, similarly to \citeA{zhang2024addressing}. However, there is some sensitivity to the training algorithm initialization for the mean flow prediction: the response pattern in the mean flow is similar, but the response magnitude can vary by $50\%$. The sensitivity of the mean fields to the random seed is not apparent in the global ocean configuration OM4.

\subsection{*}{Choice of the filter scale in the training dataset}
The filtering operator used is a Gaussian filter implemented in the package GCM-Filters \cite{grooms2021diffusion, loose2022gcm} with width $\overline{\Delta}$, chosen in relation to the coarse grid spacing $\Delta$. The filter-to-grid width ratio parameter ($\mathrm{FGR}=\overline{\Delta}/\Delta$) represents the strength of the subfilter parameterization: relatively low value of FGR ($\overline{\Delta}/\Delta=1$) in the training dataset results in a learned parameterization that has negligible effect in online simulations. On the other hand, a relatively large value ($\mathrm{FGR}=4$) results in over-energized grid-scale features. The value used here ($\mathrm{FGR}=3$) corresponds to the strongest parameterization effect without generating grid-scale noise.  Note that the optimal FGR parameter depends on the numerical and physical dissipation schemes present in the ocean model, as the ANN subfilter parameterization alone does not produce enough grid-scale dissipation. For the discussion of how to choose FGR parameter see \citeA{perezhogin2025large, perezhogin2024stable, perezhogin2023subgrid}.

\section{*}{Text S4. Online implementation and numerical stability in MOM6}

The trained parameters of the ANN subfilter model are saved to a NetCDF file and read by the numerical ocean model during initialization. The neural network inference is implemented using the Fortran module of \citeA{Sane2023}. The ANN inference takes $\approx$ 10\% of the ocean model runtime, for a neural network with one hidden layer and 20 neurons. However, the inference can be further accelerated, as we found that the inference in Python is generally faster than in Fortran. 

The implemented ANN parameterization works stably (free of NaNs in prognostic fields) in idealized Double Gyre and global ocean OM4 configurations, without any tuning, in part because the biharmonic Smagorinsky model provides the dissipation. In the idealized configuration NeverWorld2 (NW2, \citeA{marques2022neverworld2}), however, tuning is required to improve the numerical stability even when a backscatter parameterization (whether our ANN or more traditional parametrization) is used together with biharmonic Smagorinsky model; e.g., \citeA{yankovsky2024}. 
We have modified the ANN parameterization to achieve stability, without optimizing for online metrics, using a set of minimal changes. Our tuning includes attenuating the magnitude of the ANN parameterization in high-strain regions following \citeA{perezhogin2024stable} and allowing the MOM6 dynamical core to truncate velocities if they are too big. Additionally, at resolution $1/6^\circ$ in NW2, we had to reduce the time stepping interval.

\end{article}
\newpage

\begin{table}[]
\centering
\caption{Parameters of the training data and artificial neural network (ANN) model}
\label{tab:parameters}
\begin{tabular}{ll}
\toprule
\textbf{Category} & \textbf{Value} \\
\midrule
\textbf{Training Data Parameters} & \\
High-resolution data & CM2.6 \cite{griffies2015impacts}, $0.1^\circ$ ocean grid\\ 
Diagnosed features & $\overline{\sigma}_S$, $\overline{\sigma}_T$, $\overline{\omega}$, $\mathbf{S}$, $\mathbf{T}$  \\
Layer Depths (m) & $5$, $55$, $110$, $180$, $330$, $730$, $1500$, $2500$, $3500$, $4500$ \\
Horizontal grid type & Tripolar \\
Horizontal extent & All globe including polar latitudes \\
Coarse Grid Spacing, $\Delta$ (nominal) & $0.4^\circ$, $0.9^\circ$, $1.2^\circ$, $1.5^\circ$ \\
Coarse Grid Spacing, $\Delta$ (km, $60\mathrm{S}^\circ-60\mathrm{N}^\circ$) & 22-44, 50-100, 67-134, 85-167 \\
Gaussian Filter Width, $\overline{\Delta}$ & $1.2^\circ$, $2.7^\circ$, $3.6^\circ$, $4.5^\circ$; i.e., $\overline{\Delta}/\Delta=3$ \\
Training / Validation / Test Splitting (years) & $181-188$ / $194$ / $199-200$ \\
Snapshot Averaging Interval & 5 days \\
Time Seperation Between Snapshots & 1 month \\
Number of 2D Snapshots used for Training & $10 \times 4 \times 8 \times 12 = 3840$ \\
Number of Training Iterations & $16000$ (each iteration randomly selects 2D snapshot) \\
\midrule
\textbf{ANN Parameters} & \\
Input Size & $3 \times 3$ \\
ANN type & Multilayer Perceptron (MLP) \\
ANN used for offline analysis & 2 hidden layers, 32 neurons each, 2051 parameters \\
ANN used in online implementation & 1 hidden layer with 20 neurons, 623 parameters \\
Activation Function & ReLU \\
Note & Regularization is not applied during training \\
\bottomrule
\end{tabular}
\end{table}

\begin{table}[]
	\begin{center}
		\begin{tabular}{l|cccc}
			       RMSE  & $0^{\circ}\mathrm{E}$  & $15^{\circ}\mathrm{E}$ & $30^{\circ}\mathrm{E}$ & $45^{\circ}\mathrm{E}$\\ 
            \hline 
            Control($1/3^{\circ}$)     	 &  51.3 &  46.8 & 49.1 & 
               36.8  \\
               Yankovsky24($1/3^{\circ}$)     	 &  33.5 & 31.6 & 29.9 &  27.4   \\
               ZB20-Reynolds($1/3^\circ$) & $\mathbf{32.7}$ &  25.4 &  $\mathbf{26.1}$ &  $\mathbf{21.6}$ \\
               ANN($1/3^\circ$)             &  35.0 &  $\mathbf{24.3}$ &  30.9 &  28.6 \\
            \hline
		    Control($1/4^{\circ}$)     	 &  52.1  & 42.1 & 40.3 & 34.6  \\
               Yankovsky24($1/4^{\circ}$)    &  27.8  & 23.0 & 20.7 & 21.3  \\
               ZB20-Reynolds($1/4^{\circ}$)  &  $\mathbf{26.9}$  & 21.0 & 18.4 & $\mathbf{18.5}$ \\
               ANN($1/4^{\circ}$)       & $29.2$ & $\mathbf{20.0}$ & $\mathbf{16.7}$ & $19.7$ \\
            \hline
            Control($1/6^\circ$) & 42.7 &  30.7 &  31.8 &  26.7  \\
            Yankovsky24($1/6^\circ$) & 26.2 & 22.3 & 16.1 & 16.8  \\
            ZB20-Reynolds($1/6^\circ$) &  27.7 &  24.5 & 18.9 & 18.4 \\
            ANN($1/6^\circ$) & $\mathbf{23.5}$ & $\mathbf{18.8}$ & $\mathbf{13.8}$ & $\mathbf{14.6}$ 
		\end{tabular}
		\caption{Online results in idealized configuration NeverWorld2 at three coarse resolutions ($1/3^\circ$, $1/4^\circ$ and $1/6^\circ$). The root mean squared errors (RMSE) in 1000-day averaged position of interfaces over four meridional transects at longitudes $0^{\circ}\mathrm{E}$, $15^{\circ}\mathrm{E}$, $30^{\circ}\mathrm{E}$ and $45^{\circ}\mathrm{E}$. RMSE units are metres. The interfaces for Control and  ANN-parameterized runs at resolution $1/4^\circ$ at longitudes $0^{\circ}\mathrm{E}$ and $45^{\circ}\mathrm{E}$ are also shown in Figure S3. The error is computed w.r.t. $1/32^{\circ}$ model. Yankovsky24 stands for parameterization of \citeA{yankovsky2024}, ZB20-Reynolds stands for \citeA{zanna2020data} parameterization implemented and modified by \citeA{perezhogin2024stable}. Parameterizations are not retuned when resolution is changed.}
	\end{center}
\end{table}

\begin{table}[]
	\begin{center}
		\begin{tabular}{l|c}
                    & ACC transport [Sv] \\
            \hline
            $1/32^\circ$ & 235.3 \\
            \hline 
               Control($1/3^{\circ}$)     	 &  242.7  \\
               Yankovsky24($1/3^\circ$)     & $\mathbf{237.4}$ \\
               ZB20-Reynolds($1/3^\circ$)     & 230.2  \\
               ANN($1/3^\circ$)             &  241.6 \\
            \hline
		    Control($1/4^{\circ}$)     	 & 245.1  \\
               Yankovsky24($1/4^{\circ}$)    &  229.9  \\
               ZB20-Reynolds($1/4^{\circ}$)  &  225.4 \\
               ANN($1/4^{\circ}$)            & $\mathbf{236.9}$ \\
            \hline
               Control($1/6^{\circ}$)     	 & 243.3   \\
               Yankovsky24($1/6^{\circ}$)    &  $\mathbf{230.4}$   \\
               ZB20-Reynolds($1/6^{\circ}$)  &  219.5  \\
               ANN($1/4^{\circ}$)            & 228.8 \\
		\end{tabular}
		\caption{Online results in idealized configuration NeverWorld2. The ACC transport through the Drake Passage at $0^\circ \mathrm{E}$ averaged over 800 days.}
        \end{center}
\end{table}

\clearpage

\begin{figure}[h!]
\centering{\includegraphics[width=0.9\textwidth]{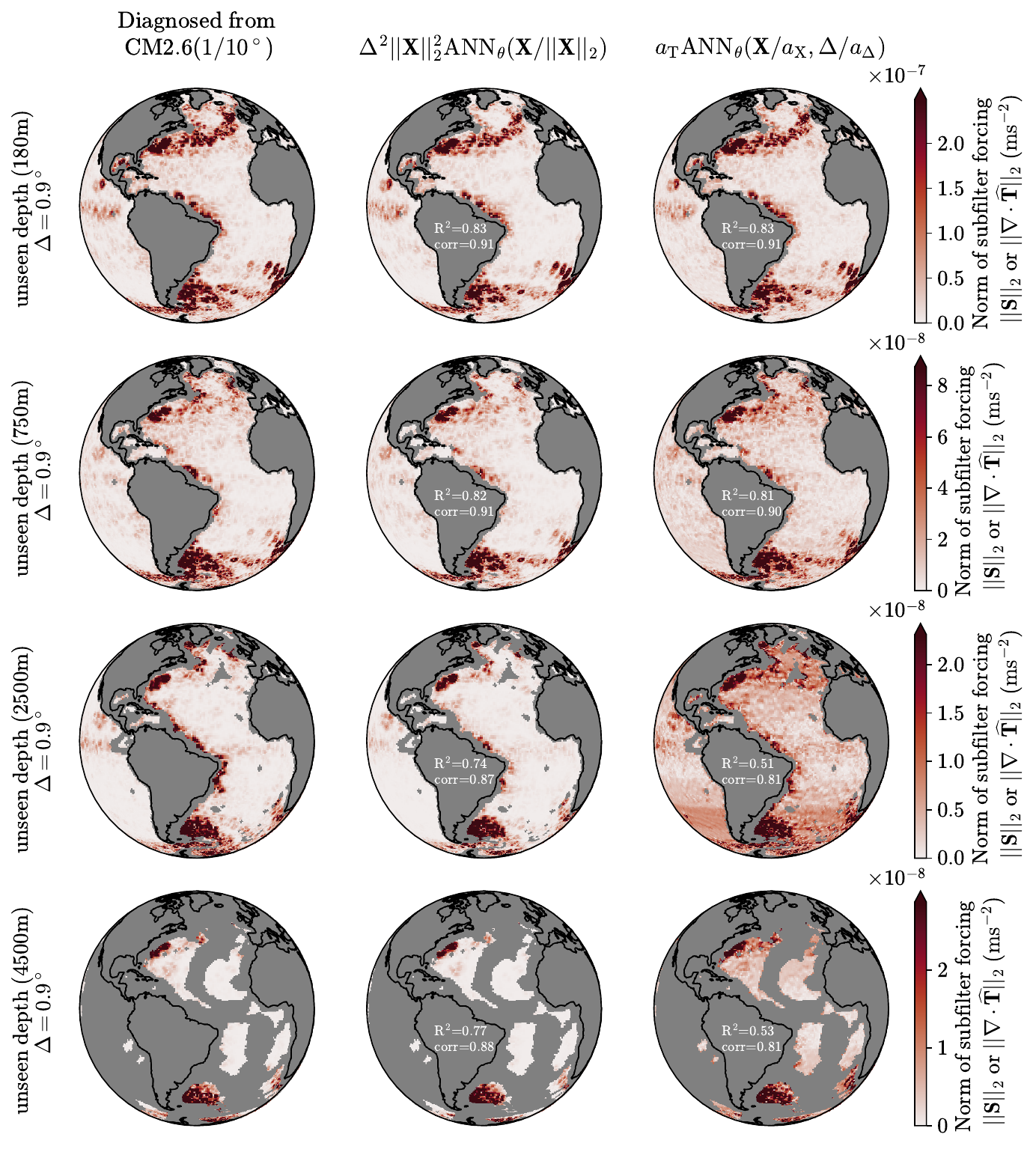}}
\caption{Extension of Figure 1 in the main text with generalization of two ANN parameterizations to multiple unseen depths, but seen resolution used for training ($0.9^\circ$).
}
\end{figure}

\begin{figure}[h!]
\centering{\includegraphics[width=0.95\textwidth]{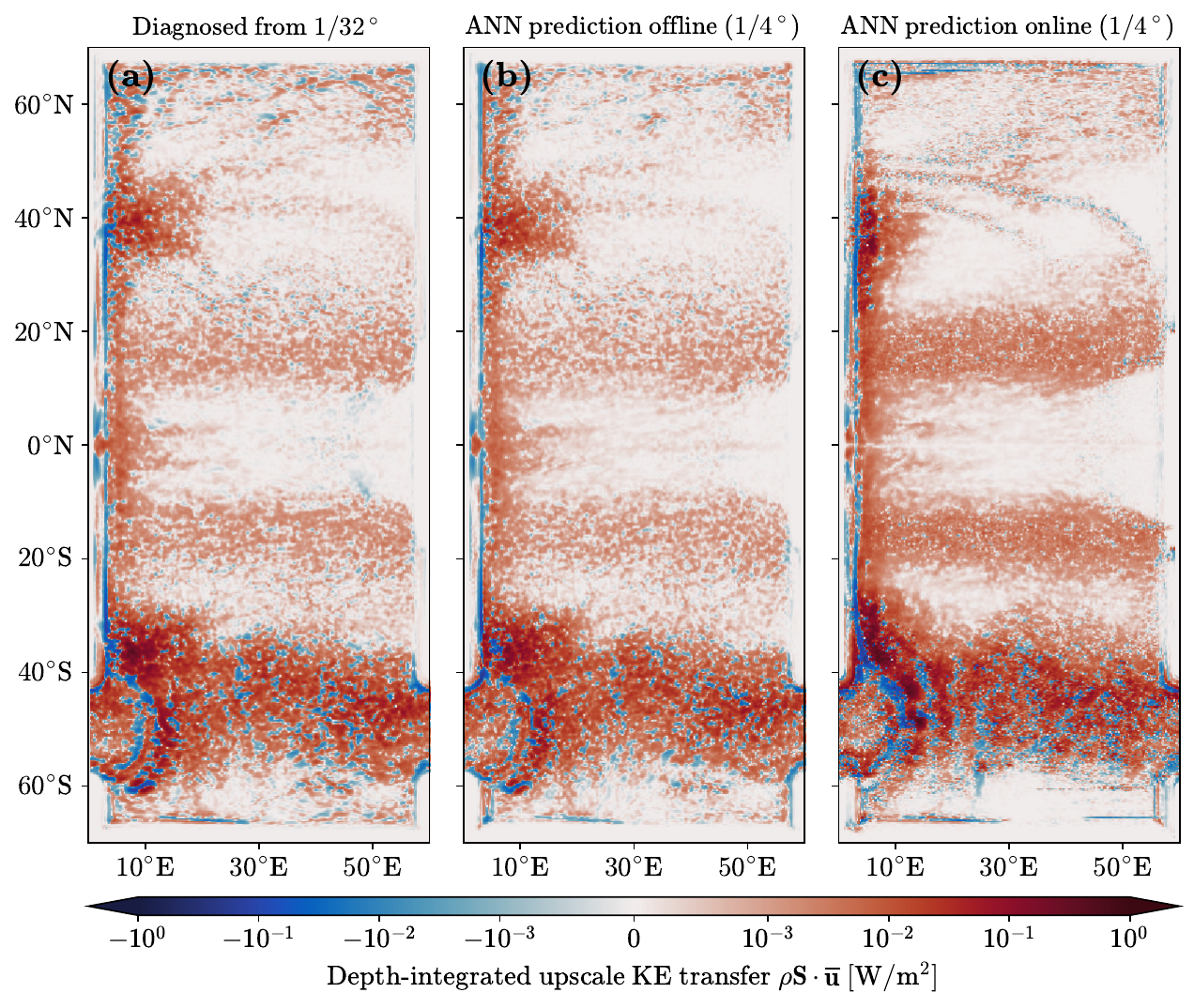}}
\caption{Upscale KE transfer (positive numbers correspond to backscatter) averaged over 800 days and integrated over depth in idealized NeverWorld2 configuration. (a) Diagnosed from high-resolution ($1/32^\circ$) simulation by filtering and coarsegraining, (b) and (c) predicted by the ANN offline and online, respectively, at coarse resolution $1/4^\circ$. The ANN was trained on global ocean data and thus generalizes well to a new configuration as seen in the accurate prediction of KE transfer offline. Prediction offline means that filtered and coarsegrained snapshots of the high-resolution model were given as inputs to the ANN. Slight degradation of prediction online is related to the difference in magnitude of small-scale velocity gradients and large-scale circulation patterns in the coarse ocean model.
}
\end{figure}

\begin{figure}[h!]
\centering{\includegraphics[width=1.0\textwidth]{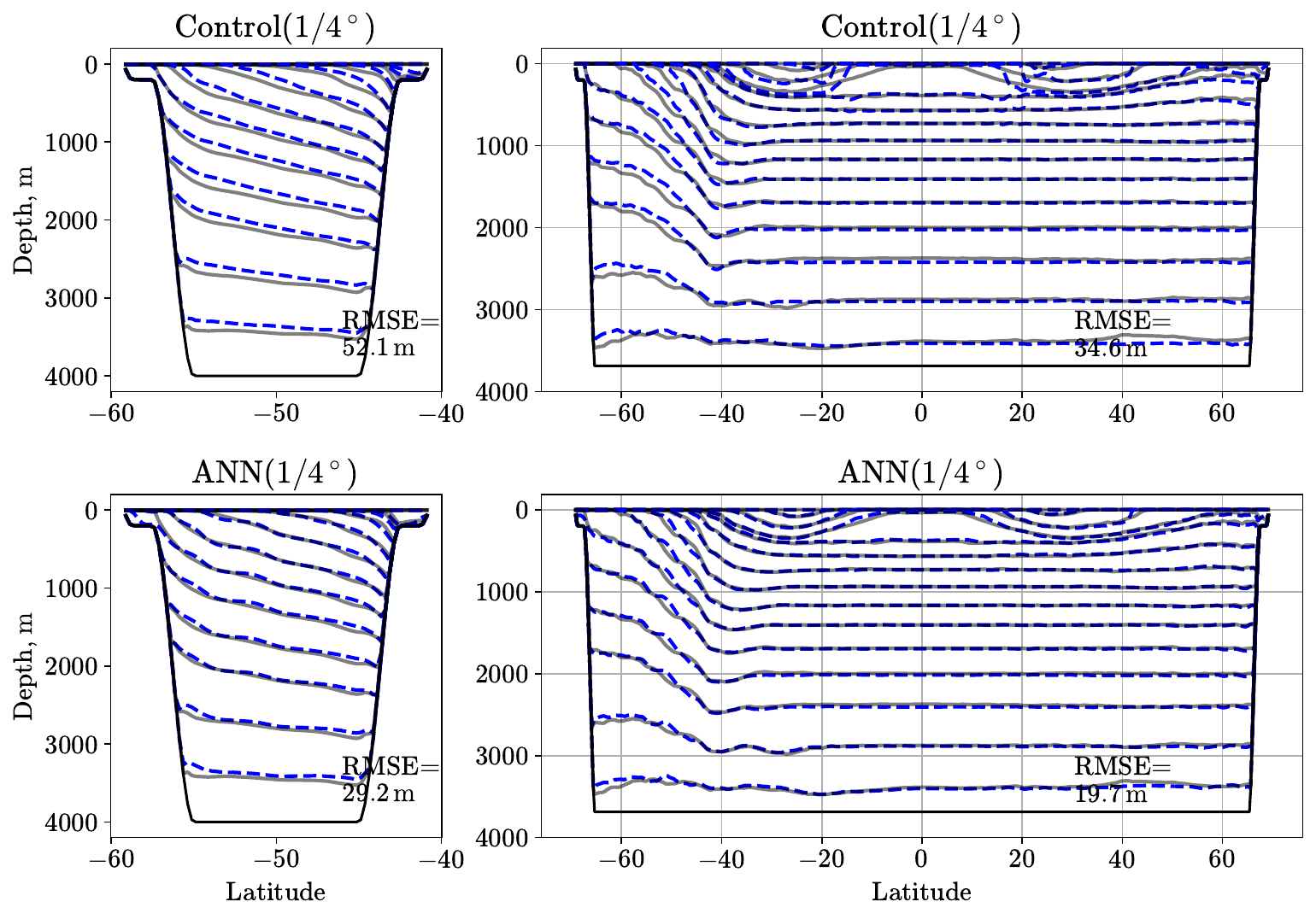}}
\caption{Online results in idealized configuration NeverWorld2. The 1000-days averaged isopycnal interfaces in the meridional transect of Drake Passage (Longitude $0^\circ \mathrm{E}$, left column) and at Longitude $45^\circ \mathrm{E}$. The blue dashed lines show the position of interfaces in the coarse-resolution ($1/4^\circ$) experiment, and the gray lines show the interfaces of the high‐resolution model $1/32^\circ$.  The root mean squared errors (RMSE) between coarse and high-resolution models are provided.
}
\end{figure}


\begin{figure}[h!]
\centering{\includegraphics[width=0.95\textwidth]{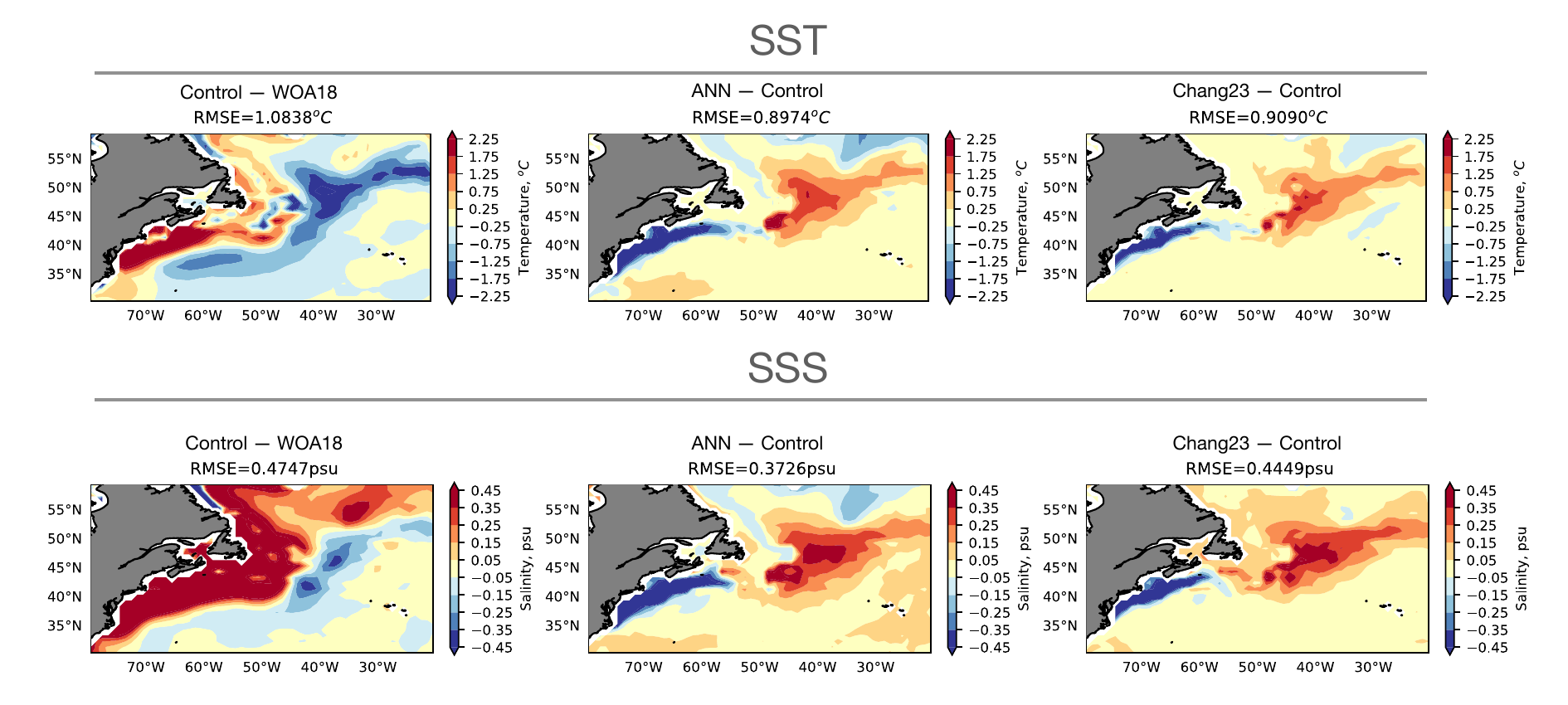}}
\caption{Online results in the global ocean-ice model OM4 \cite{adcroft2019gfdl}, North Atlantic region. Comparison of the ANN parameterization to a baseline parameterization tested in \citeA{chang2023remote}. We consider biases in sea surface temperature (SST), sea surface salinity (SSS). Model output is averaged over years 1981-2007. The observational data for SST and SSS is given by the World Ocean Atlas 2018 (WOA18, \citeA{locarnini2018world}). Root mean square errors (RMSEs) between simulations and observations are provided. 
}
\end{figure}

\begin{figure}[h!]
\centering{\includegraphics[width=0.78\textwidth]{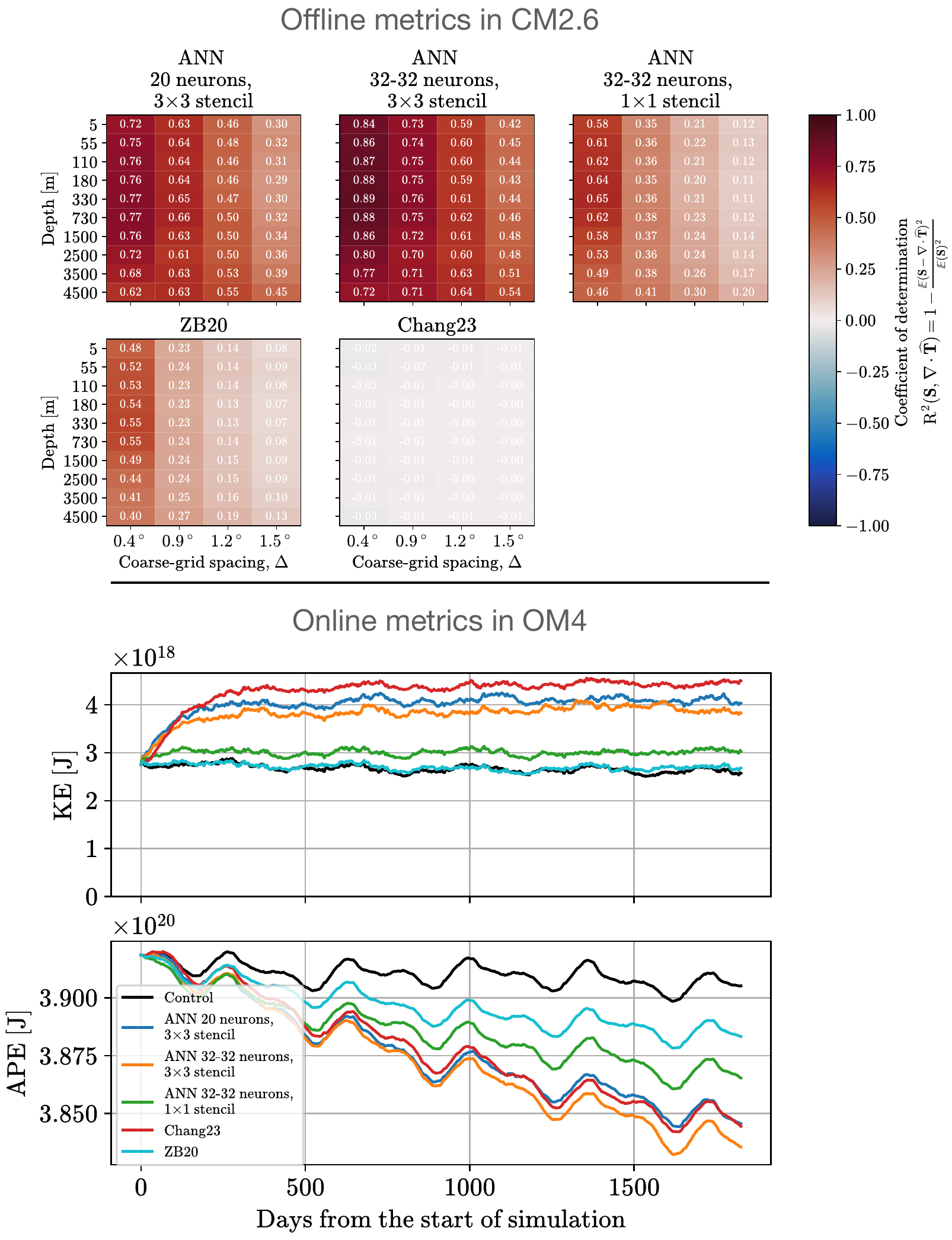}}
\caption{Upper block: Offline performance of the mesoscale eddy parameterizations on CM2.6 data. Three versions of the ANN parameterization with dimensional scaling are shown, which are different in the number of neurons used or the size of the spatial stencil. Existing parameterizations, equation-discovery model (ZB20, \citeA{zanna2020data}) and anti-viscosity model \cite{chang2023remote}, are shown for comparison. Lower block: Kinetic energy (KE) and available potential energy (APE) in short 5-year OM4 parameterized simulations.}
\end{figure}

\begin{figure}[h!]
\centering{\includegraphics[width=1.0\textwidth]{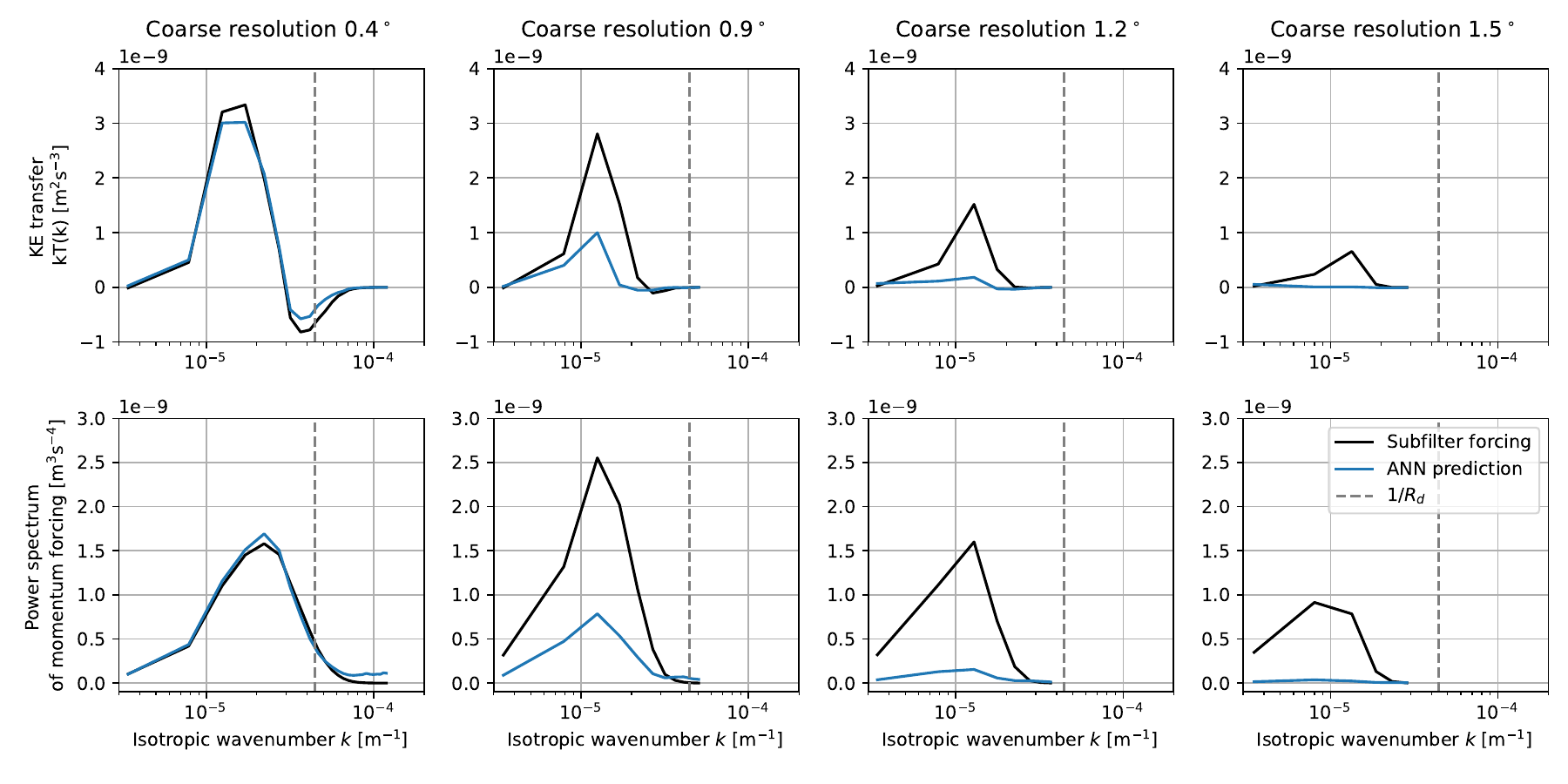}}
\caption{(Upper row) Offline kinetic energy (KE) transfer spectrum, where $T(k) = 2 \pi k Re(\mathcal{F}(\mathbf{u})^*\mathcal{F}(\mathbf{S}))$, and $\mathcal{F}$ is the 2D Fourier transform, $Re$ is the real part, and $*$ is the complex conjugate. (Lower row) power spectrum of subfilter forcing $2\pi k \mathcal{F}(\mathbf{S})^*\mathcal{F}(\mathbf{S})$. Spectra are computed in the North Atlantic region $(25-45^\circ\mathrm{N})$$\times$$(60-40^\circ\mathrm{W})$ and at depth $5$m. $R_d=22.6$km is the Rossby deformation radius in this region. Results are shown for an ANN used in online simulations.

We can identify two effects of the coarsening of the resolution on the diagnosed and predicted eddy fluxes. First, the diagnosed interscale energy transfer vanishes once the Rossby deformation radius becomes unresolved. This can be explained by the blocking of the inverse energy cascade on the scales much larger than the forcing scale (deformation radius). Second, the ANN parameterization predicts even smaller kinetic energy transfer at these coarse resolutions ($\approx 1^\circ$).

It is a subject of future studies whether we should attempt to achieve more accurate predictions at these resolutions with improved architecture of the ANN or consider alternative parameterization approaches, such as parameterizing buoyancy fluxes instead, \citeA{balwada2025design}.
}
\end{figure}


\clearpage

\bibliography{agusample}